\documentclass[a4paper,11pt]{article}
\pdfoutput=1

\usepackage{jheppub}
\usepackage[T1]{fontenc}

\title{\boldmath Coherent quantum-circuit simulation of Mueller dipole branching at small $x$: multiplicity fluctuations and DIS phenomenology}

\author[a,b,c,d]{Wei Kou}
\author[a,b,c,d,1]{Xurong Chen\note{Corresponding author}}

\affiliation[a]{Institute of Modern Physics, Chinese Academy of Sciences,
Lanzhou 730000, China}
\affiliation[b]{Southern Center for Nuclear-Science Theory (SCNT),
Institute of Modern Physics, Chinese Academy of Sciences,
Huizhou 516000, China}
\affiliation[c]{School of Nuclear Science and Technology, University of Chinese Academy of Sciences, Beijing 100049, China}
\affiliation[d]{State Key Laboratory of Heavy Ion Science and Technology, Institute of Modern Physics, Chinese Academy of Sciences, Lanzhou 730000, China}

\emailAdd{kouwei@impcas.ac.cn}
\emailAdd{xchen@impcas.ac.cn}

\abstract{
Mueller dipole branching at small $x$ is a stochastic process
in which splitting rates depend on transverse geometry.
We construct a coherent quantum circuit for a regulated
finite-state version of this process.
Poisson-clock uniformization controls the number of updates,
while orthogonal event-history registers retain the
splitting paths, including those leading to the same final
configuration. The circuit prepares transition amplitudes
from local splitting rates rather than loading a precomputed
evolved distribution.
Its statevector multiplicity probabilities reproduce the
truncated-clock Markov result to numerical precision and
agree with the full continuous-time Markov-chain benchmark
within approximately $7\times10^{-4}$.
To examine fluctuations beyond the small circuit, we
independently simulate the Mueller cascade with continuous
transverse coordinates.
At the largest rapidity studied, its normalized fourth
factorial moment is $1.81$, compared with $2.84$ for an
equal-mean Yule process.
We also determine multiplicity entropies and correlations
between dipole-size sectors.
Applying a fixed negative-binomial dipole-to-charged-particle
response reproduces the rising trend of the H1
charged-particle multiplicity entropy toward smaller $x$,
but does not fully describe the measured multiplicity
distributions.
These results demonstrate an explicit coherent encoding
of finite-state dipole branching and clarify its relation
to continuous-cascade fluctuations and DIS phenomenology.
}


\begin{document}
\maketitle
\flushbottom

\section{Introduction}
\label{sec:introduction}

At small Bjorken $x$, successive soft-gluon emissions generate
logarithmically enhanced contributions to deep-inelastic scattering (DIS)
\cite{KuraevLipatovFadin1977,Mueller1994Soft}.
In the large-$N_c$ and leading-logarithmic approximations,
Mueller's dipole picture describes each emission as the splitting
of a parent color dipole into two daughter dipoles
\cite{Mueller1994Soft,MuellerPatel1994,Mueller1995Unitarity}.
Repeated splittings produce a stochastic cascade whose evolution
depends on the transverse sizes and positions of the dipoles.
This geometrical dependence affects not only the mean dipole
multiplicity but also its fluctuations and correlations.
Although dipole evolution is closely related to small-$x$
scattering frameworks
\cite{Balitsky1996,Kovchegov1999,IancuLeonidovMcLerran2001},
the present work focuses on the regulated Mueller branching
process rather than the nonlinear evolution of scattering
amplitudes.

With finite transverse regulators and a finite configuration
space, Mueller branching can be formulated as a continuous-time
Markov chain. Its probabilities evolve according to a master
equation, while individual cascades can be generated by
Monte Carlo sampling. Multiplicity distributions and their
scaling have long been studied in high-energy collisions
\cite{KobaNielsenOlesen1972}, including recent investigations
of dipole multiplicities and entropies
\cite{Levin2024Multiplicity,KutakLokos2025,KutakPraszalowicz2025}.
The difficulty is that each splitting changes both the dipole
number and the transverse configuration, which determines
subsequent emission rates. The number of accessible configurations
therefore grows rapidly as the transverse resolution and
multiplicity cutoff increase. A further challenge arises when
representing this stochastic evolution on a quantum computer:
Markov transition probabilities cannot be directly implemented
as unitary evolution. A quantum-circuit construction must preserve
the required probabilities while keeping the evolution reversible.

Quantum computing has been applied to high-energy processes,
including parton-level dynamics, parton showers, and interference
effects
\cite{LammLawrenceYamauchi2020,NachmanProvasoliDeJongBauer2021,
BepariMalikSpannowskyWilliams2021,BauerChigusaYamazaki2024}.
Quantum simulations of gauge theories have also addressed
jet production and chiral dynamics in the Schwinger model
\cite{FlorioEtAl2023,FlorioEtAl2024,IkedaEtAl2024},
as well as DIS-related current correlators and multiparticle
processes in QCD media
\cite{IkedaKangKharzeevQian2026,BarataLiQianSalgadoSilva2026}.
Our approach addresses a different problem. Rather than
simulating real-time evolution under a QCD Hamiltonian,
we construct a coherent representation of the classical
Markov dynamics governing Mueller dipole branching.
Poisson-clock uniformization expresses the continuous-time
evolution through discrete stochastic updates. These updates
are implemented coherently using transition amplitudes
determined by local splitting rates. Orthogonal event-history
registers distinguish different branching paths, including
those reaching the same final configuration, so that their
probabilities are recovered without unwanted interference.
The construction provides an explicit quantum-circuit
realization of the finite-state process, without assuming
a quantum computational advantage.

The physical consequences of Mueller branching extend
beyond the mean multiplicity. Higher factorial moments
probe the shape of the multiplicity distribution, while
Shannon and R\'enyi entropies characterize its information
content. Correlations between different dipole-size sectors
provide additional information about the transverse structure
of the cascade. These quantities can be studied independently
using Monte Carlo simulations with continuous transverse
coordinates and compared with simpler branching models,
such as the Yule process at the same mean multiplicity.
The distinction between the finite-state circuit and the
continuous-coordinate calculation is important: the former
tests the coherent implementation of stochastic transitions,
whereas the latter examines multiplicity fluctuations in
a larger configuration space.

The relation of partonic fluctuations and entropy to measured
hadronic final states has attracted considerable interest
\cite{KovnerLublinsky2015,HagiwaraHattaXiaoYuan2018,
KharzeevLevin2017,TuKharzeevUllrich2020,
KharzeevLevin2021,HentschinskiKutak2022,
HentschinskiKutakErratum2023,HentschinskiKutakStraka2022,
HentschinskiKharzeevKutakTu2024,Kou:2022dkw}.
The H1 measurements of charged-particle multiplicity
distributions in the hadronic centre-of-mass frame,
within $0<\eta^*<4$, provide an experimental reference
\cite{H1HEPData2021}.
However, dipole multiplicities are not directly observable
charged-particle multiplicities. Their comparison requires
a phenomenological description of particle production.
Moreover, the Shannon entropy of a multiplicity distribution
is an information entropy and should not be identified
with quantum entanglement entropy without additional
physical assumptions. It is therefore necessary to examine
both the measured multiplicity distributions and their
associated entropies.

In this work, we first construct and validate a coherent
quantum circuit for regulated finite-state Mueller branching,
using a Poisson clock and orthogonal event histories.
We compare the resulting multiplicity probabilities with
an independently calculated Markov-chain benchmark.
We then investigate the continuous transverse-coordinate
cascade through high-statistics Monte Carlo simulations,
focusing on multiplicity distributions, higher factorial
moments, information entropies, and dipole-size correlations.
An equal-mean Yule process serves as a reference for
identifying differences in the fluctuation structure.
Finally, we introduce a fixed negative-binomial response
to examine how the calculated dipole statistics relate to
the charged-particle multiplicity distributions and entropies
measured by H1. This comparison tests the phenomenological
connection without identifying the finite-state circuit
with a complete simulation of the DIS final state.

The remainder of the paper is organized as follows.
Sections~\ref{sec:mueller}--\ref{sec:validation}
present the branching model, coherent-circuit construction,
and numerical validation.
Sections~\ref{sec:fluctuations} and~\ref{sec:dis}
discuss the continuous-cascade statistics and DIS comparisons.
Section~\ref{sec:conclusions} summarizes the results,
with technical details given in the appendices.

\section{Mueller dipole evolution and multiplicity observables}
\label{sec:mueller}

\subsection{Regulated Mueller branching dynamics}
\label{subsec:branching}

\begin{figure}[htbp]
\centering
\includegraphics[width=0.88\textwidth]{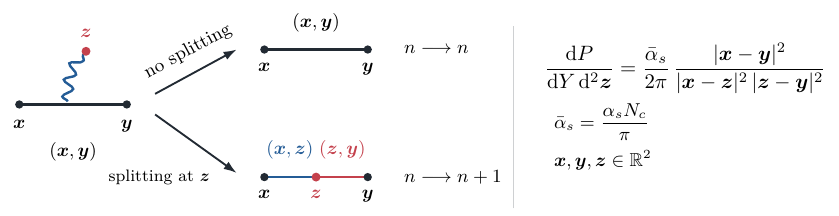}
\caption{
Elementary splitting in the Mueller color-dipole
picture. A parent dipole with transverse endpoints
$\boldsymbol{x}$ and $\boldsymbol{y}$ either remains
unchanged or splits into two daughter dipoles
following a gluon emission at $\boldsymbol{z}$.
The splitting probability is governed by
eq.~\eqref{eq:mueller_kernel}, and each accepted
splitting increases the dipole multiplicity by one.
}
\label{fig:mueller-splitting}
\end{figure}

We consider Mueller dipole branching in the large-$N_c$
and leading-logarithmic approximations, with a fixed
strong coupling \cite{Mueller1994Soft,MuellerPatel1994}.
A color dipole with transverse endpoints
$(\boldsymbol{x},\boldsymbol{y})$ and size
$r=|\boldsymbol{x}-\boldsymbol{y}|$ can emit a soft
gluon at transverse position $\boldsymbol{z}$.
In the dipole picture, this emission replaces the
parent dipole by two daughter dipoles,
$(\boldsymbol{x},\boldsymbol{z})$ and
$(\boldsymbol{z},\boldsymbol{y})$.
The splitting probability per unit evolution rapidity
$Y$ and transverse area is

\begin{equation}
\label{eq:mueller_kernel}
\frac{d\mathcal{P}_{\boldsymbol{x}\boldsymbol{y}
\rightarrow \boldsymbol{x}\boldsymbol{z},
\boldsymbol{z}\boldsymbol{y}}}
{dY\,d^2\boldsymbol{z}}
=
\frac{\bar{\alpha}_s}{2\pi}
\frac{|\boldsymbol{x}-\boldsymbol{y}|^2}
{|\boldsymbol{x}-\boldsymbol{z}|^2
 |\boldsymbol{z}-\boldsymbol{y}|^2},
\qquad
\bar{\alpha}_s=\frac{\alpha_s N_c}{\pi},
\end{equation}
where $\alpha_s$ is the strong coupling and $N_c$
is the number of colors. Each splitting increases the
dipole multiplicity by one. Because the daughter-dipole
geometry determines subsequent emission rates, the
cascade is a configuration-dependent branching process,
rather than a pure-birth process specified by the
dipole number alone
\cite{Salam1995Multiplicity,MuellerSalam1996Fluctuations}.
The elementary splitting is illustrated in
figure~\ref{fig:mueller-splitting}.

The splitting kernel is singular when the emission
position approaches either endpoint of the parent
dipole. We therefore introduce a regulated transverse
region $\Omega_{\mathrm{reg}}(\boldsymbol{x},
\boldsymbol{y})$ to obtain finite emission rates
\cite{Salam1995Multiplicity,Salam1997OEDIPUS}.
Specifically, the emission position is restricted to
a square domain $[-L,L]^2$, while both daughter-dipole
lengths must lie between the ultraviolet cutoff
$\rho$ and the infrared scale $R_{\mathrm{IR}}$.
The regulated splitting rate of a single dipole is

\begin{equation}
\label{eq:regulated_rate}
\gamma(\boldsymbol{x},\boldsymbol{y})
=
\frac{\bar{\alpha}_s}{2\pi}
\int_{\Omega_{\mathrm{reg}}
(\boldsymbol{x},\boldsymbol{y})}
d^2\boldsymbol{z}\,
\frac{|\boldsymbol{x}-\boldsymbol{y}|^2}
{|\boldsymbol{x}-\boldsymbol{z}|^2
 |\boldsymbol{z}-\boldsymbol{y}|^2}.
\end{equation}

For a configuration
$\mathcal{C}=\{(\boldsymbol{x}_i,\boldsymbol{y}_i)\}_{i=1}^{n}$,
the total rate of the uncapped continuous-coordinate
cascade is

\begin{equation}
\label{eq:total_branching_rate}
\Gamma_{\mathrm{cont}}(\mathcal{C})
=
\sum_{i=1}^{N(\mathcal{C})}
\gamma(\boldsymbol{x}_i,\boldsymbol{y}_i),
\end{equation}
where $N(\mathcal{C})=n$ denotes the number of dipoles
in the configuration. The waiting interval in rapidity
until the next splitting is exponentially distributed
with rate $\Gamma_{\mathrm{cont}}(\mathcal{C})$.
The emitting dipole is selected according to its
relative splitting rate, and the emission position
is distributed according to the regulated kernel.
The resulting daughter dipoles then participate
in subsequent emissions \cite{Salam1997OEDIPUS}.

We use two numerical representations of this regulated
branching process. For the continuous-coordinate
Monte Carlo calculation, an event-driven Poisson-thinning
algorithm samples emission positions continuously
within the allowed transverse region. A grid of
upper bounds is used to generate candidate emissions,
but accepted transverse positions are not projected
onto that grid. For the finite-state Markov-chain
and quantum-circuit calculations, the transverse
region is instead divided into emission cells.
The transition rate of each cell is obtained by
numerically integrating the regulated kernel over
that cell, while an allowed representative position
specifies the daughter-dipole geometry.
This produces a finite set of accessible configurations
and transition channels.

The finite-state calculation additionally imposes
a maximum dipole multiplicity $n_{\max}$.
Configurations satisfying
$N(\mathcal{C})=n_{\max}$ are absorbing, with
vanishing outgoing rate $\Gamma(\mathcal{C})=0$.
This cutoff is a numerical restriction and does
not represent gluon saturation or recombination.
Both numerical descriptions retain the same
regulated Mueller splitting kernel, but their
different treatments of transverse geometry
and multiplicity truncation mean that they are
not identical Markov processes.
The regulator parameters, finite-state construction,
and cutoff tests are specified in
appendix~\ref{app:regularization}.
The present model describes the branching sector
only; nonlinear BK/JIMWLK evolution and the
scattering-amplitude dynamics are not included.

\subsection{Markov description and multiplicity observables}
\label{subsec:markov}

We first describe the regulated finite-state cascade as a
continuous-time Markov chain. Let $\mathcal S$ denote the set of
accessible dipole configurations and $p_{\mathcal C}(Y)$ the
probability of configuration $\mathcal C\in\mathcal S$ at
evolution rapidity $Y$. Using a column probability vector
$\boldsymbol p(Y)$, the master equation and its solution are
\begin{equation}
\label{eq:markov_master}
\frac{d\boldsymbol p(Y)}{dY}=Q\boldsymbol p(Y),
\qquad
\boldsymbol p(Y)=e^{QY}\boldsymbol p(0).
\end{equation}
Here $Q$ is the Markov generator. Each allowed splitting
channel contributes its rate to the transition between the
corresponding initial and final configurations. If several
channels lead to the same final configuration, their rates
are added. Accordingly,
$Q_{\mathcal C',\mathcal C}\geq0$ for
$\mathcal C'\neq\mathcal C$,
$Q_{\mathcal C,\mathcal C}=-\Gamma(\mathcal C)$, and
$\sum_{\mathcal C'}Q_{\mathcal C',\mathcal C}=0$.
The last condition ensures probability conservation.
At the multiplicity cap $n_{\max}$, outgoing splitting
rates vanish, making the corresponding configurations
absorbing. The explicit construction of $Q$ is given in
appendix~\ref{app:finite-generator}.

The probability of finding $n$ dipoles is obtained by
summing over all configurations with that multiplicity:
\begin{equation}
\label{eq:dipole_multiplicity}
P_n(Y)=
\sum_{\substack{\mathcal C\in\mathcal S\\
N(\mathcal C)=n}}
p_{\mathcal C}(Y),
\qquad
P_n(0)=\delta_{n1}.
\end{equation}
Here $N(\mathcal C)$ is the number of dipoles in
configuration $\mathcal C$. For the finite-state chain,
$1\leq n\leq n_{\max}$ and
$\sum_nP_n(Y)=1$, including the probability accumulated
at the absorbing boundary. The continuous-coordinate
Monte Carlo calculation provides an independent estimate
of the multiplicity distribution. In that calculation,
events exceeding the imposed multiplicity cap are recorded
separately, and normalized statistical observables are
evaluated for the retained event sample. The treatment of
these events and the associated uncertainties is detailed
in appendix~\ref{app:mc-estimators}.

To characterize fluctuations beyond the mean dipole
number
\cite{Salam1995Multiplicity,Levin2021Factorial,
MuellerSalam1996Fluctuations,LiouMuellerMunier2017},
we introduce the probability-generating function
\begin{equation}
\label{eq:multiplicity_generating_function}
\mathcal G(u,Y)=\sum_{n\geq1}P_n(Y)u^n,
\qquad
\langle n\rangle_Y=
\left.\frac{\partial\mathcal G}{\partial u}
\right|_{u=1},
\end{equation}
where $u$ is an auxiliary generating variable.
Successive derivatives of $\mathcal G$ give the factorial
moments of the multiplicity distribution. We define
the normalized factorial moments as
\begin{equation}
\label{eq:factorial_moments}
F_q(Y)=
\frac{\langle(n)_q\rangle_Y}
{\langle n\rangle_Y^q},
\qquad
(n)_q=n(n-1)\cdots(n-q+1).
\end{equation}
In particular, we calculate $F_2$, $F_3$, and $F_4$
to examine fluctuations beyond the mean multiplicity.
These quantities are compared with a Yule pure-birth
process whose mean multiplicity is matched to the
Mueller result separately at each rapidity
\cite{KutakLokos2025}.
The comparison tests differences in the shapes of the
multiplicity distributions without attributing them
simply to different mean multiplicities.
The Yule expressions, statistical estimators, and
multiplicity-cap dependence are discussed in
appendix~\ref{app:statistics}.

We also characterize the multiplicity distribution
through its Shannon and R\'enyi entropies
\cite{Levin2024Multiplicity,KutakLokos2025,
HentschinskiJungKutak2026}:
\begin{equation}
\label{eq:multiplicity_entropies}
\begin{aligned}
S_1(Y)
&=-\sum_{n\geq1}P_n(Y)\ln P_n(Y),\\
S_\alpha(Y)
&=\frac{1}{1-\alpha}
\ln\!\left[
\sum_{n\geq1}P_n^\alpha(Y)
\right],
\qquad \alpha>0,\quad\alpha\neq1.
\end{aligned}
\end{equation}
Natural logarithms are used, with $0\ln0\equiv0$.
We report the Shannon entropy $S_1$ and the R\'enyi
entropies $S_2$ and $S_3$. While the factorial moments
emphasize different orders of multiplicity fluctuations,
these entropies characterize how probability is
distributed among multiplicity classes.
They are classical information entropies defined
from $P_n(Y)$, and should not be identified with
quantum entanglement entropy
\cite{HagiwaraHattaXiaoYuan2018,
HentschinskiEtAl2026CompleteDipole}.

Finally, to examine how branching fluctuations depend
on transverse geometry, we divide dipoles into two
size sectors at a threshold $r_{\mathrm{cut}}$
\cite{Salam1995Multiplicity,MuellerSalam1996Fluctuations}.
We denote the numbers of small and large dipoles by
$n_s$ and $n_l$, corresponding to
$r<r_{\mathrm{cut}}$ and $r\geq r_{\mathrm{cut}}$,
respectively, so that $n=n_s+n_l$.
Their joint distribution $P(n_s,n_l;Y)$ contains
information that is lost when only the total
multiplicity $P_n(Y)$ is considered.
We quantify their correlation using the covariance
\begin{equation}
\label{eq:size_covariance}
C_{sl}(Y)=
\langle n_sn_l\rangle_Y
-\langle n_s\rangle_Y\langle n_l\rangle_Y.
\end{equation}
Because fluctuations in the total dipole number can
themselves induce covariance between the two sectors,
a nonzero $C_{sl}$ does not necessarily imply an
additional size-dependent correlation mechanism.
We therefore compare the measured covariance with
a conditional-binomial reference that preserves the
total-multiplicity distribution $P_n(Y)$ while
randomly assigning dipoles to the two size sectors.
The reference and its interpretation are described
in appendix~\ref{app:joint-size}.

\section{Coherent quantum-circuit representation}
\label{sec:quantum-circuit}

\subsection{Uniformization of the branching process}
\label{subsec:uniformization}

We begin with the finite-state Markov generator $Q$
defined in section~\ref{subsec:markov}. Its continuous
evolution in rapidity $Y$ is given by
$\boldsymbol p(Y)=e^{QY}\boldsymbol p(0)$.
To represent this evolution through discrete stochastic
updates, we use Poisson-clock uniformization
\cite{Jensen1953,GrossMiller1984}.

We choose a common clock rate $\Lambda>0$ satisfying
$\Lambda\geq\max_{\mathcal C\in\mathcal S}
\Gamma(\mathcal C)$, where $\Gamma(\mathcal C)$
is the total outgoing splitting rate of configuration
$\mathcal C$. The continuous-time evolution can then
be written exactly as

\begin{equation}
\label{eq:uniformization}
\begin{aligned}
T &= I+\frac{Q}{\Lambda},\\
w_k(Y) &= e^{-\Lambda Y}
\frac{(\Lambda Y)^k}{k!},\\
\boldsymbol p(Y)
&=\sum_{k=0}^{\infty}
w_k(Y)T^k\boldsymbol p(0).
\end{aligned}
\end{equation}
Here $T$ is a column-stochastic transition matrix,
and $w_k(Y)$ is the Poisson probability of $k$ clock
updates during the rapidity interval $Y$.
For a given configuration $\mathcal C$, each update
produces an allowed splitting through channel $a$
with probability $\gamma_a(\mathcal C)/\Lambda$,
or leaves the configuration unchanged with probability
$1-\Gamma(\mathcal C)/\Lambda$.
These null updates account for the configuration
dependence of the physical splitting rates.
Consequently, $k$ counts clock updates rather than
the number of emitted gluons.

For a finite circuit, we retain at most $K$ clock
updates. The omitted Poisson probability is

\begin{equation}
\label{eq:clock_truncation}
\varepsilon_K(Y)
=
1-e^{-\Lambda Y}
\sum_{k=0}^{K}
\frac{(\Lambda Y)^k}{k!}.
\end{equation}

In our coherent implementation, the retained clock
sectors and the unresolved tail are prepared in
a single quantum register:

\begin{equation}
\label{eq:coherent_clock_state}
|\omega_K(Y)\rangle
=
\sum_{k=0}^{K}\sqrt{w_k(Y)}\,|k\rangle
+
\sqrt{\varepsilon_K(Y)}\,|K+1\rangle.
\end{equation}
The orthogonal state $|K+1\rangle$ labels all
clock histories beyond the retained depth.
The clock state is normalized, and its different
sectors coherently control the corresponding
numbers of stochastic updates. No intermediate
measurement or classical sampling of $k$ is
required. The controlled updates are constructed
from local splitting probabilities, as described
in the following subsection.

The multiplicity probabilities associated with
the resolved sectors have total weight
$1-\varepsilon_K(Y)$. For comparison with the
normalized finite-state Markov distribution,
we use the conditional result

\begin{equation}
\label{eq:finite_clock_conditional}
\boldsymbol p^{(K)}(Y)
=
\frac{1}{1-\varepsilon_K(Y)}
\sum_{k=0}^{K}
w_k(Y)T^k\boldsymbol p(0).
\end{equation}
The difference between this conditional distribution
and the exact Markov result is bounded by
$\varepsilon_K(Y)$ in total variation distance.
This error concerns only the finite Poisson-clock
depth for a fixed generator $Q$. The transverse
discretization and absorbing multiplicity cutoff
introduce separate approximations.
The normalization and error bound are discussed
further in appendix~\ref{app:clock}.

\subsection{Coherent encoding of branching histories}
\label{subsec:coherent-encoding}

The stochastic transition matrix $T$ cannot itself act
as a unitary operator on quantum amplitudes. We therefore
encode each update through the square roots of its
transition probabilities, while retaining the information
needed to distinguish different branching histories.
For a dipole configuration $\mathcal C$, the probability
of a physical splitting at one Poisson-clock update can
be represented by a rotation angle $\theta(\mathcal C)$,
defined through
\begin{equation}
\label{eq:branch_rotation}
\sin^2\theta(\mathcal C)
=
\frac{\Gamma(\mathcal C)}{\Lambda}.
\end{equation}
A splitting through channel $a$ occurs with conditional
probability $\gamma_a(\mathcal C)/\Gamma(\mathcal C)$
when $\Gamma(\mathcal C)>0$.
Thus the complete one-step event probabilities are
\begin{equation}
\label{eq:local_event_probabilities}
\pi(e|\mathcal C)=
\begin{cases}
1-\Gamma(\mathcal C)/\Lambda,
    & e=0,\\[2mm]
\gamma_e(\mathcal C)/\Lambda,
    & e\in\mathcal A(\mathcal C),
\end{cases}
\end{equation}
where $e=0$ denotes a null update and
$\mathcal A(\mathcal C)$ is the set of allowed
splitting channels. For an absorbing configuration,
$\pi(0|\mathcal C)=1$.

Starting from the initial configuration $\mathcal C_0$,
a sequence of $k$ clock updates defines an event history
$h=(e_1,\ldots,e_k)$. The configuration after each
update is determined recursively by the previous
configuration and the recorded event.
Writing $\mathcal C_t(h)$ for the configuration after
$t$ updates, the probability of a particular history is
\[
\mathbb P(h|k)
=
\prod_{t=1}^{k}
\pi\!\left(e_t\middle|
\mathcal C_{t-1}(h)\right).
\]
Both physical splittings and null updates are retained
in the history. Distinct histories are assigned
orthogonal computational-basis states, even if they
produce the same final dipole configuration.

Combining these histories with the coherent Poisson
clock of section~\ref{subsec:uniformization}, we obtain
the finite-depth state
\begin{equation}
\label{eq:coherent_history_state}
\begin{aligned}
|\Psi_K(Y)\rangle
={}&
\sum_{k=0}^{K}\sqrt{w_k(Y)}
\sum_{h\in\mathcal H_k}
\sqrt{\mathbb P(h|k)}
\,|k\rangle|h\rangle\\
&+
\sqrt{\varepsilon_K(Y)}
\,|K+1\rangle|\boldsymbol{0}\rangle_h ,
\end{aligned}
\end{equation}
where $\mathcal H_k$ contains the allowed event histories
of length $k$. The history registers have a fixed
maximum depth $K$, with unused registers initialized
to zero. The clock value distinguishes inactive
registers from null updates, and the last term retains
the unresolved Poisson tail.

In the implemented circuit, the event-history registers
also determine the current dipole configuration.
Consequently, no separate configuration register is
required for state preparation from the fixed initial
dipole. Controlled rotations prepare the event
probabilities $\pi(e|\mathcal C)$ according to the
configuration reconstructed from the preceding history.
The circuit uses local transition probabilities rather
than the precomputed evolved distribution
$\boldsymbol p(Y)$. The gate construction is described
in section~\ref{subsec:circuit-readout}.

The orthogonality of the history states ensures that
different paths reaching the same configuration
contribute their probabilities without interference.
For a fixed number $k$ of updates, tracing over
the history labels gives
\begin{equation}
\label{eq:coherent_probability}
\begin{aligned}
p_{\mathcal C}^{(k)}
&=
\sum_{\substack{h\in\mathcal H_k\\
\mathcal C_k(h)=\mathcal C}}
\left|\langle h|\psi_k\rangle\right|^2\\
&=
\left[T^k\boldsymbol p(0)\right]_{\mathcal C},
\end{aligned}
\end{equation}
where
$|\psi_k\rangle=
\sum_{h\in\mathcal H_k}
\sqrt{\mathbb P(h|k)}\,|h\rangle$.
Together with the Poisson weights, this recovers
the truncated Markov evolution of
section~\ref{subsec:uniformization}.
The construction is unitary on the enlarged register
space, but its amplitudes encode classical branching
probabilities rather than quantum interference
between QCD emission amplitudes.
A general isometric formulation and its relation
to the history encoding are given in
appendix~\ref{app:history-isometry}.

\subsection{Circuit implementation and readout}
\label{subsec:circuit-readout}

\begin{figure}[htbp]
\centering
\includegraphics[width=0.95\textwidth]{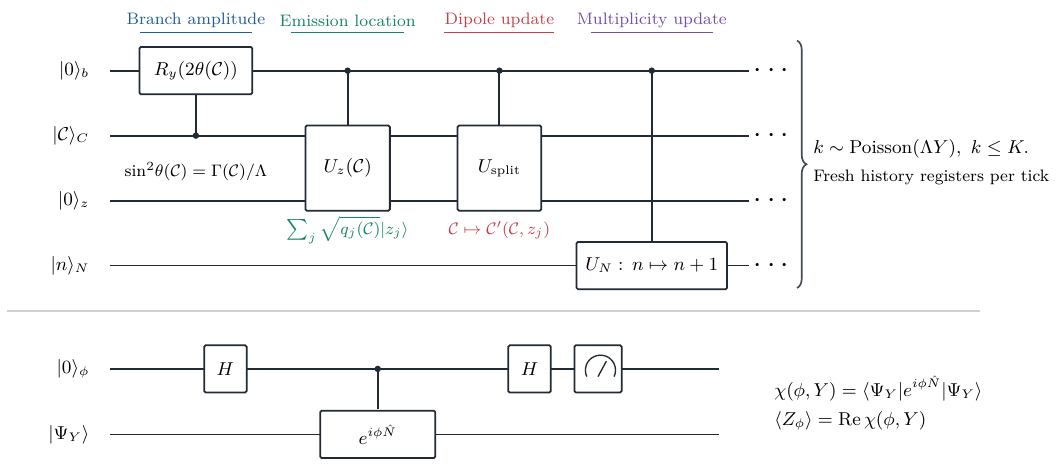}
\caption{
Schematic quantum-circuit representation of regulated
Mueller dipole branching and multiplicity readout.
The upper panel illustrates configuration-dependent
branching, emission-channel selection, and the retention
of orthogonal histories. These operations describe the
logical branching structure rather than the literal
register layout of the implemented circuit, which uses
a coherent Poisson clock and event-history registers.
The lower panel illustrates ancilla-assisted readout
of the multiplicity characteristic function.
The characteristic-function curves reported in this
work are obtained directly from simulated statevectors;
independent ancilla measurements are separately checked
using Qiskit Aer.
}
\label{fig:coherent-circuit}
\end{figure}

We implement the coherent state of
eq.~\eqref{eq:coherent_history_state} using a Poisson-clock
register and a sequence of event-history registers.
Figure~\ref{fig:coherent-circuit} illustrates the
corresponding operations and the readout of multiplicity
observables. The upper panel shows the branching operations
schematically; the implemented circuit encodes configurations
implicitly through their event histories, rather than
maintaining separate configuration and multiplicity registers.

The circuit begins by preparing the clock amplitudes
$\sqrt{w_k(Y)}$ for $0\leq k\leq K$, together with
the orthogonal tail state carrying weight
$\varepsilon_K(Y)$. For each active clock step, a new
event register is prepared conditionally on the clock
value and the preceding history. The current dipole
configuration is reconstructed from that history, and
the corresponding local probabilities
$\pi(e|\mathcal C)$ of
eq.~\eqref{eq:local_event_probabilities}
determine the rotation angles.
We use a sequence of controlled $R_y$ rotations to
prepare the binary encoding of the event distribution,
following the general principle of recursive
probability-amplitude state preparation
\cite{GroverRudolph2002,Mottonen2005StatePrep}.
For a binary prefix with probability weights
$M_0$ and $M_1$ in its two branches, the rotation
angle is $2\arcsin\sqrt{M_1/(M_0+M_1)}$.
The controls select the appropriate clock sector
and previously recorded events. No intermediate
measurement is performed, and null updates remain
part of the recorded history.

In this implementation, a zero event code represents
a null update, while a nonzero code identifies an
allowed dipole splitting. For a history
$h=(e_1,\ldots,e_k)$, the final dipole multiplicity
is therefore
\begin{equation}
\label{eq:history_multiplicity}
N(h)=1+\sum_{t=1}^{k}\mathbf{1}_{e_t\neq0},
\end{equation}
where $\mathbf{1}_{e_t\neq0}$ equals one for a
physical splitting and zero for a null update.
Measuring the clock and history registers, reconstructing
$N(h)$, and summing over histories with the same
multiplicity gives the resolved probabilities
\begin{equation}
\label{eq:history_multiplicity_readout}
P_{n,\mathrm{res}}^{(K)}(Y)
=
\sum_{k=0}^{K}w_k(Y)
\sum_{\substack{h\in\mathcal H_k\\N(h)=n}}
\mathbb P(h|k).
\end{equation}
These probabilities have total weight
$1-\varepsilon_K(Y)$. The conditional distribution
$P_n^{(K)}(Y)$ follows by dividing by this weight,
as in eq.~\eqref{eq:finite_clock_conditional}.

Beyond the multiplicity probabilities, we consider
the characteristic function
\begin{equation}
\label{eq:multiplicity_characteristic}
\chi(\phi,Y)
=
\sum_{n\geq1}P_n(Y)e^{i\phi n}
=
\langle\Psi_Y|
e^{i\phi\hat N}
|\Psi_Y\rangle,
\end{equation}
where $\phi$ is the counting phase and
$\hat N$ is the dipole-number observable.
In the history encoding, $\hat N$ is evaluated
through $N(h)$ rather than a separate multiplicity
register. The expectation-value expression applies
to the normalized encoding of the complete
finite-state Markov distribution.

For the finite clock, the resolved characteristic
function is
\begin{equation}
\label{eq:finite_clock_characteristic}
\chi_{\mathrm{res}}^{(K)}(\phi,Y)
=
\sum_n P_{n,\mathrm{res}}^{(K)}(Y)e^{i\phi n},
\qquad
\chi^{(K)}(\phi,Y)
=
\frac{\chi_{\mathrm{res}}^{(K)}(\phi,Y)}
{1-\varepsilon_K(Y)}.
\end{equation}
An ancilla-assisted controlled-phase circuit can
access the real and imaginary parts through
measurements of the ancilla Pauli operators
$X$ and $Y$. In our implementation, the unresolved
clock-tail sector is assigned zero counting phase.
The raw ancilla signal is consequently
\begin{equation}
\label{eq:ancilla_characteristic_signal}
\langle X\rangle+i\langle Y\rangle
=
\varepsilon_K(Y)
+\chi_{\mathrm{res}}^{(K)}(\phi,Y).
\end{equation}
Subtracting the known tail contribution and applying
the finite-clock normalization yields
$\chi^{(K)}(\phi,Y)$.

We evaluate the characteristic-function curves
directly from the coherent statevectors and compare
them with the matched Markov-chain results in
appendix~\ref{app:resources-fcs}.
Independent Qiskit Aer measurements of the ancilla
in the $X$ and $Y$ bases provide an additional
sampling check at a selected counting phase.
These are simulated measurements, not results
from physical quantum hardware.

The circuit thus provides a direct test of the
coherent implementation of the finite-state
Markov process. Its transverse resolution,
absorbing multiplicity cutoff, and retained clock
depth limit the dynamics represented in this
calculation. The associated history-register
requirements and logical circuit resources are
discussed in appendix~\ref{app:resources-fcs}.
The continuous-coordinate cascade used for
higher-multiplicity physics is studied separately
in section~\ref{sec:fluctuations}.

\section{Quantum simulation and numerical validation}
\label{sec:validation}

We validate the coherent history circuit against the
continuous-time Markov chain (CTMC) constructed from the
same finite-state generator $Q$. Both calculations use
identical splitting rates, transverse discretization,
initial conditions, and absorbing multiplicity cutoff.
The CTMC probabilities are obtained independently from
the Markov evolution, without using the quantum circuit.
This comparison tests whether the circuit correctly
implements the finite-state stochastic dynamics. It does
not, by itself, establish convergence to the
continuous-coordinate Mueller cascade.

The quantum benchmark uses a $2\times2$ transverse
emission grid and a maximum dipole multiplicity
$n_{\max}=3$. The resulting finite model contains
37 accessible configurations. We prepare the coherent
Poisson clock and event-history registers using the
controlled rotations described in
section~\ref{subsec:circuit-readout}, and evolve the
resulting Qiskit circuit with a statevector simulator.
The final dipole multiplicity is reconstructed from the
measured clock sector and event history, rather than
from an independent multiplicity register.

We first check the implementation against the
finite-depth uniformization formula. Denoting by
$P_{n,\mathrm{res}}^{\mathrm{QS}}(Y)$ the resolved
probabilities extracted from the circuit, we define
\begin{equation}
\label{eq:qs_uniformization_error}
\delta_{\mathrm{impl}}(Y)
=
\max_n
\left|
P_{n,\mathrm{res}}^{\mathrm{QS}}(Y)
-
\sum_{k=0}^{K}w_k(Y)
\left[T^k\boldsymbol p(0)\right]_n
\right|,
\end{equation}
where $[\boldsymbol v]_n$ denotes the sum of
configuration probabilities with multiplicity $n$.
For all tested statevector points,
$\delta_{\mathrm{impl}}$ is at the level of
$10^{-12}$. This confirms that the constructed circuit
reproduces the intended truncated stochastic evolution
to numerical precision.

\begin{figure}[htbp]
\centering
\includegraphics[width=0.95\textwidth]{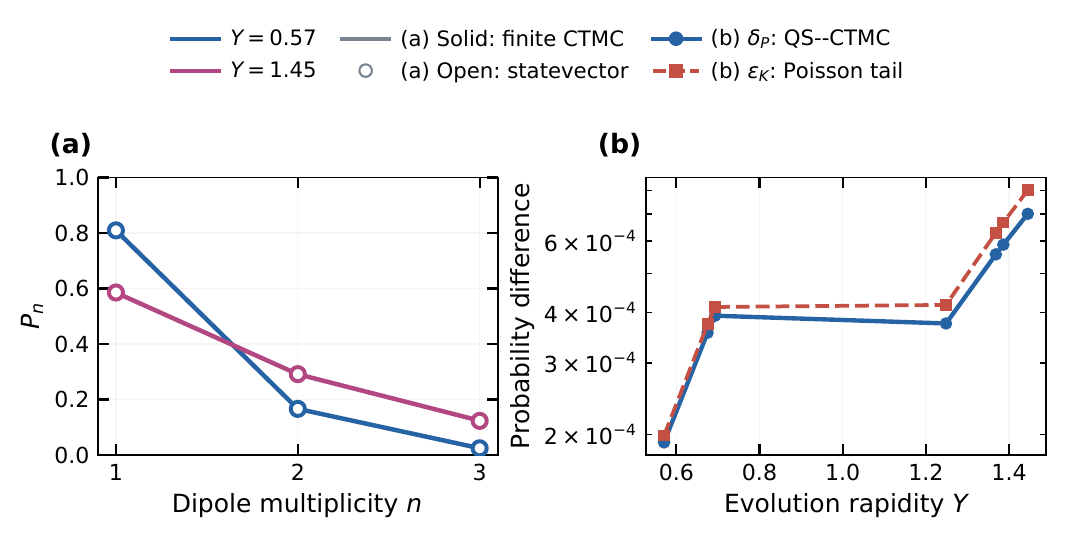}
\caption{
Validation of the coherent statevector simulation
against the matched finite-state continuous-time
Markov chain (CTMC).
(a) Dipole multiplicity distributions for $n=1,2,3$
at $Y=0.57$ and $Y=1.45$. Solid curves show the
CTMC probabilities, while open symbols represent
the conditional coherent-statevector results.
(b) Maximum absolute probability difference
$\delta_P(Y)$ (blue series) and the omitted
Poisson-clock probability $\varepsilon_K(Y)$
(red dashed series).
The observed differences remain below the
clock-tail bound for the same finite-state generator.
No finite-shot sampling or quantum-hardware errors
are included in these statevector comparisons.
}
\label{fig:qs-ctmc}
\end{figure}

We then compare the normalized finite-clock result
with the full CTMC distribution. The corresponding
error is
\begin{equation}
\label{eq:qs_ctmc_error}
\delta_P(Y)
=
\max_n
\left|
P_n^{\mathrm{QS}}(Y)
-
P_n^{\mathrm{CTMC}}(Y)
\right|,
\end{equation}
where $P_n^{\mathrm{QS}}(Y)$ denotes the conditional
finite-clock distribution of
eq.~\eqref{eq:finite_clock_conditional}.
Figure~\ref{fig:qs-ctmc}(a) compares the probabilities
for $n=1,2,3$ at $Y=0.57$ and $Y=1.45$.
The statevector and CTMC results are nearly
indistinguishable on the scale of the plot.

As shown in figure~\ref{fig:qs-ctmc}(b), the maximum
difference remains below the omitted Poisson-clock
weight $\varepsilon_K(Y)$ throughout the tested
rapidity range. The largest observed discrepancy is
approximately $7.0\times10^{-4}$, compared with a
clock-tail probability of approximately
$8.0\times10^{-4}$ at the same rapidity.
Since the circuit already agrees with finite-depth
uniformization to numerical precision, the larger
difference from CTMC is consistent with the finite
clock truncation, rather than an error in preparing
the branching amplitudes. The bound derived in
appendix~\ref{app:clock} applies to the same
finite-state generator and multiplicity cutoff.

The coherent simulations use a rapidity-dependent
clock depth chosen to keep the omitted Poisson
weight below the prescribed tolerance. At the
nonzero rapidities explicitly simulated with
Qiskit statevectors, the retained depths are
$K=3$ and $K=4$, requiring 15 and 19 logical
qubits, respectively, in the implemented
history encoding. The corresponding circuit
construction uses 148 and 333 controlled
rotations before decomposition into a
hardware-native gate set.
These counts characterize the constructed
logical circuits and should not be interpreted
as optimized hardware gate counts or depths.

The resource estimate shown in
figure~\ref{fig:app-cap-resources}(b) extends
to approximately 23 logical qubits when a
larger clock depth is required. This is an
encoding-based estimate, not an additional
23-qubit statevector simulation.
The increase reflects the need to preserve
an additional event-history register for
each retained clock update. Further resource
details are given in
appendix~\ref{app:resources-fcs}.

The finite-state circuit also has limitations
that are distinct from clock truncation.
The transverse grid replaces continuous
emission positions by a finite set of
transition channels, and the absorbing
boundary at $n_{\max}=3$ suppresses further
branching once that multiplicity is reached.
As illustrated in
figure~\ref{fig:app-cap-resources}(a),
the multiplicity entropy is sensitive to
a restrictive cap, whereas increasing the
cap gives more stable results over the
tested rapidities. These effects are
properties of the finite model and are
not controlled by reducing
$\varepsilon_K(Y)$ alone.

The present quantum calculation therefore
establishes the numerical correctness of
the coherent history encoding for a
specified regulated finite-state process.
It does not demonstrate quantum computational
advantage or provide a converged
continuous-coordinate Mueller simulation.
To investigate multiplicity fluctuations
beyond the small circuit benchmark, we
next turn to an independent Monte Carlo
calculation that retains continuous
transverse emission coordinates and
allows substantially larger dipole
multiplicities.

\section{Multiplicity fluctuations and transverse correlations}
\label{sec:fluctuations}

\subsection{Multiplicity distributions}
\label{subsec:multiplicity-distributions}

We now turn to the regulated Mueller cascade with
continuous transverse coordinates. Unlike the finite-state
quantum-circuit benchmark of section~\ref{sec:validation},
this calculation retains the transverse positions of
the daughter dipoles after each splitting and follows
the resulting configuration-dependent evolution.
We generate $10^5$ Monte Carlo trajectories with a
maximum dipole multiplicity $n_{\max}=32$.
The numerical procedure and statistical treatment
are detailed in appendix~\ref{app:mc-estimators}.

Figure~\ref{fig:mueller-multiplicity}(a) shows the dipole
multiplicity distributions $P_n(Y)$ at three representative
rapidities, $Y=0.69$, $1.39$, and $2.14$.
Starting from a single dipole, successive splittings
populate configurations with increasing multiplicity.
Accordingly, the probability of retaining only the
initial dipole decreases with $Y$, while the
distribution extends toward larger $n$.
Because the emission rate depends on the transverse
configuration, these probabilities reflect the
accumulated branching histories rather than a
multiplicity-only birth rate.

The rapidity dependence is displayed more fully in
figure~\ref{fig:mueller-multiplicity}(b).
The initially concentrated distribution progressively
broadens, although low-multiplicity configurations
remain important throughout the studied range.
The logarithmic probability scale makes the
less populated multiplicity sectors visible.
These sectors are particularly relevant for
higher-order factorial moments, which weight
large multiplicities more strongly than the
mean. Their statistical reliability is therefore
examined separately in the following subsection.

The multiplicity cap is treated as a numerical
sampling limit rather than a physical saturation
mechanism. Trajectories that exceed the cap are
recorded as overflow events and are not added
to the $n=n_{\max}$ bin. No overflow events
are observed in the present $10^5$-trajectory
sample over the evaluated rapidity range.
Thus, for these data, the observed multiplicity
probabilities coincide with those conditioned
on no overflow. This does not imply that
overflow has strictly zero probability.
The plotted error bars are 95\% Wilson
confidence intervals for the sampled
multiplicity probabilities. Only populated
bins are displayed on the logarithmic scale;
empty sampled bins should not be interpreted
as forbidden multiplicities.

\begin{figure}[htbp]
\centering
\includegraphics[width=0.95\textwidth]{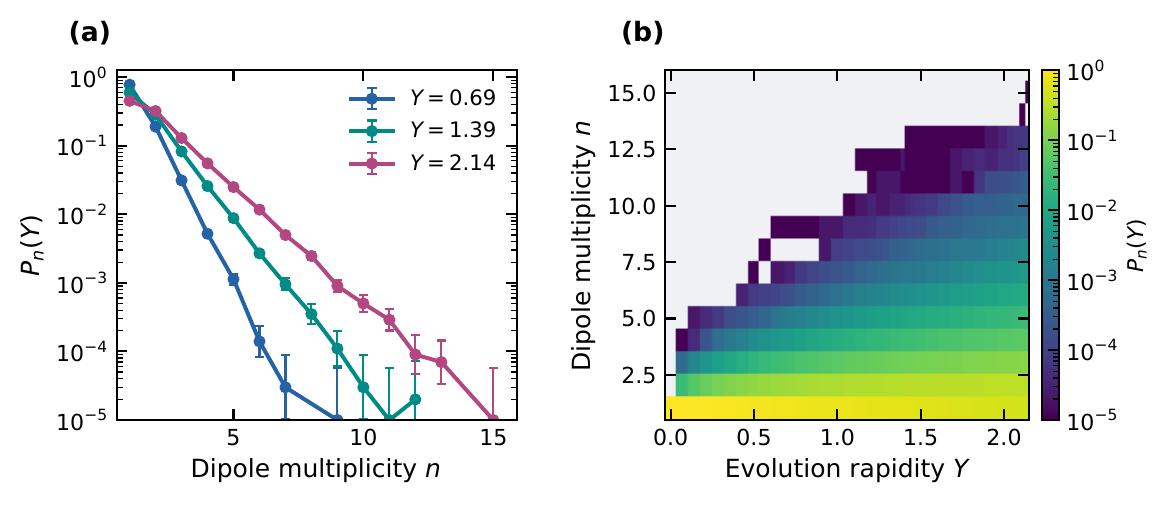}
\caption{
Dipole multiplicity distributions from the regulated
continuous-coordinate Mueller Monte Carlo cascade.
(a) $P_n(Y)$ at $Y=0.69$, $1.39$, and $2.14$ on a
logarithmic scale; error bars show 95\% Wilson
confidence intervals for the sampled probabilities.
(b) Evolution of $P_n(Y)$ with rapidity, with probability
indicated by the logarithmic color scale.
Only populated multiplicity bins are shown for
clarity; the simulation uses $n_{\max}=32$ and
records overflow separately.
}
\label{fig:mueller-multiplicity}
\end{figure}

These distributions provide the common statistical
basis for the higher factorial moments and
multiplicity entropies discussed below.
The transverse information retained in each
cascade also allows us to examine correlations
between different dipole-size sectors, which
cannot be determined from $P_n(Y)$ alone.

\subsection{Higher-order fluctuations and multiplicity entropies}
\label{subsec:higher-order-statistics}

\begin{figure}[htbp]
\centering
\includegraphics[width=0.95\textwidth]{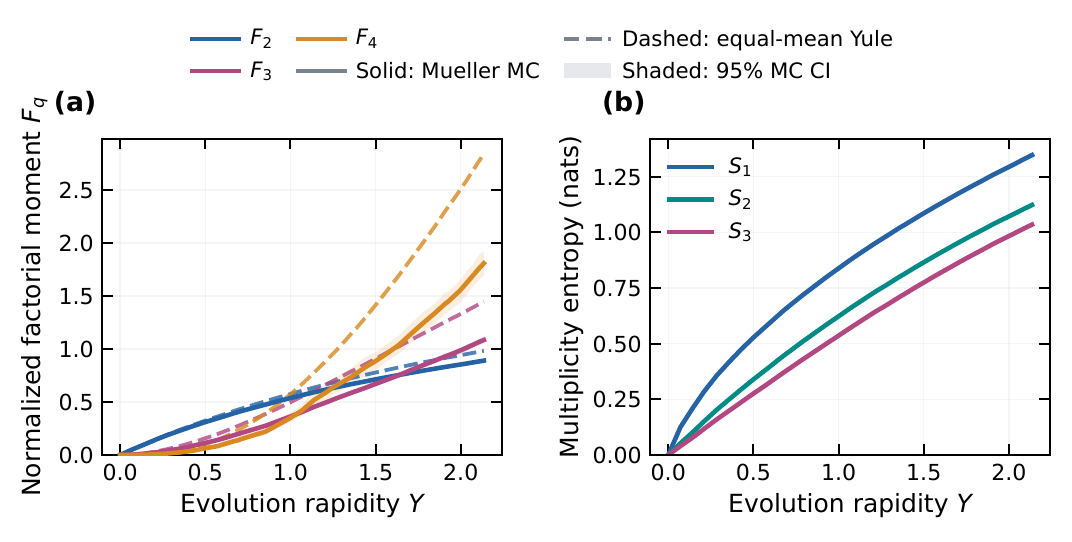}
\caption{
Higher-order multiplicity fluctuations and
information entropies in the regulated Mueller
dipole cascade.
(a) Normalized factorial moments $F_2$, $F_3$,
and $F_4$. Solid curves show the continuous-coordinate
Mueller Monte Carlo results, while dashed curves
show the analytic Yule reference matched to the
Mueller mean independently at each rapidity.
(b) Shannon entropy $S_1$ and R\'enyi entropies
$S_2$ and $S_3$ of the Mueller multiplicity
distribution, expressed in nats.
Shaded bands indicate the corresponding 95\%
Monte Carlo bootstrap confidence intervals.
}
\label{fig:mueller-moments-entropy}
\end{figure}

The multiplicity distributions discussed above contain
information beyond their mean values. To quantify these
fluctuations, we calculate the normalized factorial
moments $F_2$, $F_3$, and $F_4$ defined in
eq.~\eqref{eq:factorial_moments}.
Figure~\ref{fig:mueller-moments-entropy}(a) compares
the continuous-coordinate Mueller results with an
equal-mean Yule process. Unlike Mueller branching,
the Yule process has a splitting rate determined
only by the number of dipoles. At each rapidity,
its mean is matched to the Mueller mean, allowing
us to compare fluctuations without differences
in the average multiplicity. This is a pointwise
analytic reference, not a fit of a single Yule
splitting rate to the entire rapidity evolution.

The factorial moments increase as the cascade
evolves, but the Mueller and Yule results show
increasing differences at higher orders.
At the largest rapidity studied,
$Y\simeq2.14$, the mean dipole multiplicity is
$\langle n\rangle=1.9651$.
The Mueller cascade gives
$F_4=1.809$, with a 95\% bootstrap confidence
interval of $[1.694,1.923]$, whereas the
equal-mean Yule reference gives $F_4=2.843$.
The separation is also visible in $F_2$ and
$F_3$, although it is more pronounced for
the fourth-order moment.

This comparison demonstrates that the mean
multiplicity does not determine the higher-order
fluctuation structure. In Mueller branching,
the splitting rates change with the evolving
transverse dipole configuration, whereas the
Yule reference depends only on the total
multiplicity. The observed differences are
therefore consistent with the effects of
configuration-dependent branching within
the regulated cascade. They do not, by
themselves, isolate a unique transverse
correlation mechanism.

Higher factorial moments give increasing
weight to rare large-multiplicity events,
making their statistical precision and
sensitivity to the multiplicity cap important.
The confidence intervals in
figure~\ref{fig:mueller-moments-entropy}(a)
are obtained by bootstrap resampling of
the Monte Carlo multiplicity counts.
The treatment of overflow and the
multiplicity-cap checks are described in
appendices~\ref{app:mc-estimators}
and~\ref{app:yule-cap}, respectively.
The quoted results apply to the regulated
model and the rapidity range studied here.

The same multiplicity distributions also determine
the Shannon and R\'enyi entropies shown in
figure~\ref{fig:mueller-moments-entropy}(b).
All three entropies increase over the studied
rapidity range as the probability distribution
spreads across more multiplicity classes.
At $Y\simeq2.14$, we obtain
$S_1=1.347$, $S_2=1.123$, and $S_3=1.035$,
in natural units.
Their ordering, $S_1\geq S_2\geq S_3$,
follows from the general properties of
R\'enyi entropies. Higher R\'enyi orders
give greater relative importance to the
most probable multiplicity classes.

The factorial moments and information entropies
thus provide complementary descriptions of
the cascade. The former are particularly
sensitive to the high-multiplicity tail,
while the latter characterize the overall
distribution of probability among
multiplicity classes.
Neither quantity alone identifies how
the transverse geometry contributes to
the observed fluctuations. We therefore
examine the joint multiplicities of
different dipole-size sectors in the
following subsection.

\subsection{Transverse-size correlations}
\label{subsec:transverse-correlations}

The multiplicity distribution $P_n(Y)$ does not retain
information about the transverse sizes of individual
dipoles. To examine how the branching dynamics affects
the size composition of the cascade, we divide the
dipoles into small and large sectors using the
threshold $r_{\mathrm{cut}}=0.50$, in the transverse
coordinate units of the model. The corresponding
multiplicities are $n_s$ and $n_l$, with
$n=n_s+n_l$, as defined in
section~\ref{subsec:markov}.

\begin{figure}[htbp]
\centering
\includegraphics[width=0.95\textwidth]{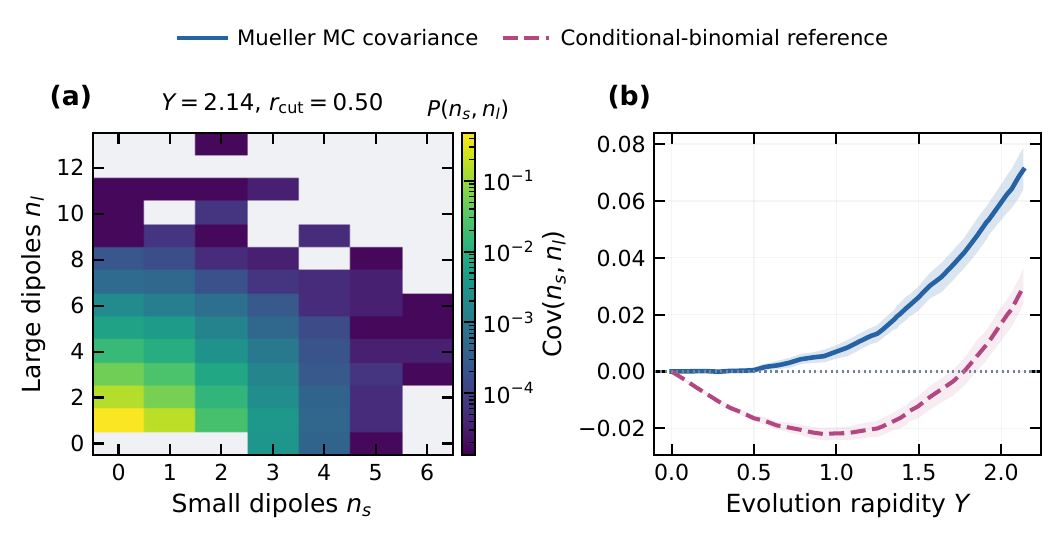}
\caption{
Transverse-size-resolved multiplicity correlations
in the regulated Mueller cascade.
(a) Joint distribution $P(n_s,n_l;Y)$ at
$Y=2.14$ and $r_{\mathrm{cut}}=0.50$,
with probability indicated on a logarithmic scale.
(b) Mueller covariance $C_{sl}(Y)$ compared with
the fixed-fraction conditional-binomial reference,
which preserves the measured total-multiplicity
distribution and the mean small-dipole fraction
at each rapidity.
Shaded bands, where available, indicate the
corresponding 95\% Monte Carlo confidence intervals.
}
\label{fig:size-correlations}
\end{figure}

Figure~\ref{fig:size-correlations}(a) shows the joint
distribution $P(n_s,n_l;Y)$ at $Y\simeq2.14$.
The distribution is concentrated at relatively low
multiplicities, but contains configurations with
different numbers of small and large dipoles.
Unlike the total multiplicity distribution,
$P(n_s,n_l;Y)$ retains information about how the
dipoles are distributed between the two size sectors.

We characterize this joint distribution through
the covariance $C_{sl}(Y)$ defined in
eq.~\eqref{eq:size_covariance}.
Figure~\ref{fig:size-correlations}(b) compares the
Mueller covariance with a conditional-binomial
reference. In this reference, the measured
total-multiplicity distribution $P_n(Y)$ is
preserved, while each dipole is independently
assigned to the small-size sector with probability
$p_s(Y)=\langle n_s\rangle_Y/\langle n\rangle_Y$.
The resulting covariance is
\begin{equation}
\label{eq:fixed_fraction_covariance}
C_{sl}^{\mathrm{bin}}(Y)
=
p_s(Y)[1-p_s(Y)]
\left[
\operatorname{Var}(n)_Y-\langle n\rangle_Y
\right].
\end{equation}

This reference accounts for correlations that can
arise from fluctuations in the total number of
dipoles, even when the size labels are assigned
independently at fixed $n$. The Mueller covariance
becomes positive and grows over the higher-rapidity
part of the studied interval, while the fixed-fraction
reference is substantially smaller. Their difference
shows that a binomial partition with a single
size-sector fraction does not reproduce the
observed joint distribution.

The fixed-fraction reference does not account for
changes in the average size composition with the
total multiplicity. We therefore perform a more
restrictive check using the same joint-distribution
data. For each total multiplicity $n$, we define
\begin{equation}
\label{eq:conditional_size_fraction}
p_s(n,Y)
=
\frac{\langle n_s\mid n,Y\rangle}{n}.
\end{equation}
The refined reference preserves both $P_n(Y)$
and the measured conditional mean
$\langle n_s\mid n,Y\rangle$, but assumes that
the number of small dipoles at fixed $n$ follows
a binomial distribution with probability
$p_s(n,Y)$. In particular, it retains the
deterministic size category of the initial
single-dipole state.

We evaluate this additional diagnostic using
$6\times10^4$ continuous-coordinate cascade
trajectories. At $Y=2.1355$, the observed covariance
is $C_{sl}=0.07096$, compared with
$C_{sl}^{\mathrm{bin},n}=0.02975$ for the
multiplicity-dependent reference. Their difference
is
\begin{equation}
\label{eq:excess_size_covariance}
\Delta C_{sl}
=
C_{sl}-C_{sl}^{\mathrm{bin},n}
=
0.04120,
\end{equation}
with a 95\% bootstrap confidence interval
of $[0.03748,0.04505]$.
Thus, allowing the mean size fraction to vary
with $n$ reduces the discrepancy but does not
eliminate it.

Because $n_l=n-n_s$, the covariance between
$n_s$ and $n_l$ at fixed $n$ is necessarily
nonpositive. The positive excess relative to
the refined reference therefore indicates
smaller fluctuations of the size composition
than expected from independent binomial
assignments at fixed multiplicity. It should
not be interpreted as a positive covariance
at fixed $n$, nor does it uniquely identify
the microscopic origin of the deviation.
The two reference constructions and their
statistical interpretation are detailed in
appendix~\ref{app:joint-size}.

These results show that the transverse-size
composition contains statistical information
beyond the total dipole multiplicity.
The observed deviations from the conditional
binomial references are consistent with the
configuration-dependent nature of Mueller
branching, while their detailed dynamical
origin requires additional observables.

\section{Connection to DIS multiplicity measurements}
\label{sec:dis}

\subsection{Shannon entropy at small $x$}
\label{subsec:dis-entropy}

We now examine how the multiplicity fluctuations of
the Mueller cascade are reflected in DIS final-state
observables. We use the H1 measurements of charged-particle
multiplicities in the current hemisphere,
$0<\eta^*<4$, of the hadronic centre-of-mass (HCM)
frame \cite{H1HEPData2021}. The measured charged-particle
multiplicity $N_{\mathrm{ch}}$ is not identical to the
partonic dipole number $n$. A comparison therefore
requires a prescription for relating the two
distributions.

For each $Q^2$ interval $b$, we introduce a conditional
charged-particle response
$R_b(N_{\mathrm{ch}}\mid n;\boldsymbol{\vartheta}_b)$.
The predicted charged-particle multiplicity distribution is
\begin{equation}
\label{eq:charged_response}
P_b^{\mathrm{ch}}(N_{\mathrm{ch}};Y)
=
\sum_{n\geq1}
R_b(N_{\mathrm{ch}}\mid n;\boldsymbol{\vartheta}_b)
P_n(Y).
\end{equation}
Here $P_n(Y)$ is obtained from the regulated Mueller
cascade, while $R_b$ represents the phenomenological
conversion from dipole number to charged-particle
multiplicity. We use a negative-binomial response
in which each dipole contributes an independent source.
For $n$ dipoles, its mean and shape parameters are
$\mu_b(n)=n\mu_b^{(0)}$ and
$\kappa_b(n)=nr_b^{(0)}$, respectively.

The two single-source response parameters are fixed
separately in each $Q^2$ interval using the
charged-particle mean and variance at a reference
point corresponding to $Y=0$. Because the initial
cascade contains one dipole, this calibration
determines the single-source response directly.
Once fixed, the same response is used at all
smaller $x$ values in that $Q^2$ interval.
No additional fit to the multiplicity entropy
or readjustment of the response at individual
$x$ points is performed. The negative-binomial
form, calibration relations, and fixed parameter
values are given in
appendix~\ref{app:nbd-response}.

We relate the evolution rapidity to Bjorken $x$
through the phenomenological assignment
$Y_b(x)=\ln(x_{\mathrm{ref},b}/x)$, where
$x_{\mathrm{ref},b}$ is the reference point in
the corresponding $Q^2$ interval.
The four intervals each contain four measured
$x$ points. Thus, four points serve to fix the
response parameters, while the remaining twelve
points test the predicted $x$ dependence without
further response fitting. The reference points
constrain the model and should not be counted
as independent tests of the evolution. The
rapidity assignment uses representative
experimental kinematics rather than an explicit
integration over the full acceptance of each bin.

From the resulting charged-particle distribution,
we calculate the Shannon entropy
\begin{equation}
\label{eq:charged_entropy}
S_{1,b}^{\mathrm{ch}}(x)
=
-\sum_{N_{\mathrm{ch}}=0}^{\infty}
P_b^{\mathrm{ch}}(N_{\mathrm{ch}};Y_b(x))
\ln P_b^{\mathrm{ch}}(N_{\mathrm{ch}};Y_b(x)).
\end{equation}
This is the information entropy of the predicted
charged-particle multiplicity distribution. It
depends on both the partonic branching dynamics
and the fixed phenomenological response, and
is not identified with quantum entanglement
entropy.

\begin{figure}[htbp]
\centering
\includegraphics[width=0.95\textwidth]{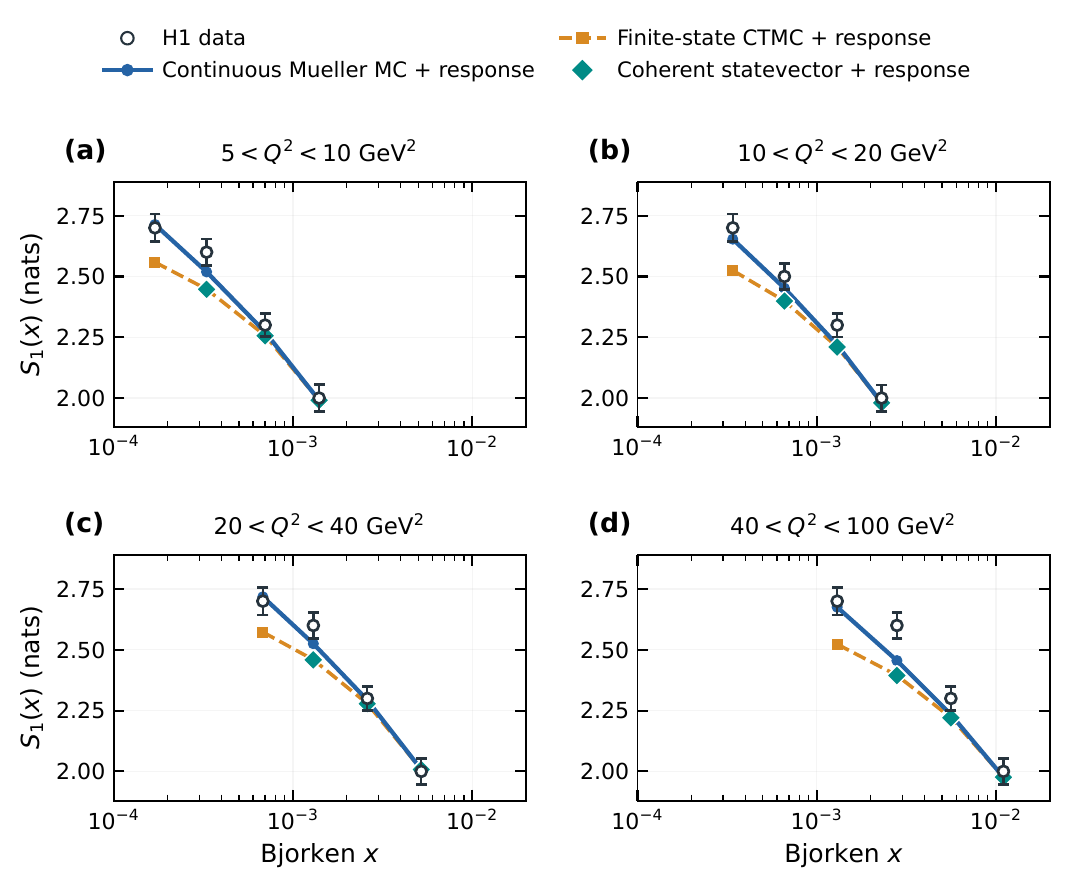}
\caption{
Charged-particle multiplicity Shannon entropy in the HCM
current hemisphere ($0<\eta^*<4$) as a function of Bjorken $x$,
compared with H1 measurements
\cite{H1HEPData2021}.
The four panels correspond to
(a) $5<Q^2<10~\mathrm{GeV}^2$,
(b) $10<Q^2<20~\mathrm{GeV}^2$,
(c) $20<Q^2<40~\mathrm{GeV}^2$, and
(d) $40<Q^2<100~\mathrm{GeV}^2$.
Open circles denote the H1 hadron entropy extracted
from measured charged-particle multiplicities, while solid lines
show the continuous Mueller Monte Carlo results
after the charged-particle response.
Dashed lines represent the finite-state CTMC
predictions, and diamond symbols indicate the
corresponding coherent statevector results.
The phenomenological response is held fixed
within each $Q^2$ interval.
}
\label{fig:h1-entropy}
\end{figure}

Figure~\ref{fig:h1-entropy} compares the
calculated entropies with the H1 measurements
in four photon-virtuality intervals:
$5<Q^2<10$, $10<Q^2<20$, $20<Q^2<40$, and
$40<Q^2<100~\mathrm{GeV}^2$.
The measured charged-particle entropy increases
as $x$ decreases in all four intervals.
The continuous-coordinate Mueller calculation,
after applying the fixed response, reproduces
this overall increasing trend.
However, quantitative differences remain,
and the agreement is not uniform across
the measured points or $Q^2$ intervals.
The finite-state CTMC and coherent statevector
results are also displayed where their more
restricted rapidity ranges permit a comparison.
They provide a separate check of the finite-state
implementation and should not be interpreted
as continuous-coordinate predictions.

The comparison is subject to uncertainties beyond
the Monte Carlo sampling error. Variations of the
transverse regulators, with the response held fixed,
produce the spread shown in
figure~\ref{fig:app-h1-regulators}.
This spread describes sensitivity to the chosen
regulated cascade and is not a statistical
confidence interval. These variations do not
systematically remove the differences from H1.
Moreover, the reference points are constrained
by the response calibration, even though their
entropies are not fitted directly.

The entropy comparison therefore provides a
phenomenological test of how the $x$ dependence
of partonic multiplicity fluctuations is transmitted
through a fixed charged-particle response.
Its qualitative agreement with the observed
entropy growth does not establish that the same
model accurately reproduces the complete
charged-particle multiplicity distribution.
We examine the distribution shapes in the
following subsection.

\subsection{Charged-particle multiplicity distributions}
\label{subsec:dis-multiplicity}

The Shannon entropy summarizes the spread of a
multiplicity distribution but does not uniquely
determine its shape. Distributions with similar
entropies may differ in their low-multiplicity
regions and high-multiplicity tails.
We therefore examine the charged-particle
multiplicity distributions measured by H1
in the HCM current hemisphere,
$0<\eta^*<4$, using the same fixed
negative-binomial response introduced in
section~\ref{subsec:dis-entropy}.

To compare with the published experimental
distributions, we account for the finite
widths of the charged-particle multiplicity
bins. For a bin $B_{bj}$ of width
$\Delta N_{bj}$ in the $Q^2$ interval $b$,
the predicted probability density per unit
charged-particle multiplicity is
\begin{equation}
\label{eq:h1_bin_density}
D_{bj}^{\mathrm{th}}(Y)
=
\frac{1}{\Delta N_{bj}}
\sum_{N_{\mathrm{ch}}\in B_{bj}}
P_b^{\mathrm{ch}}(N_{\mathrm{ch}};Y).
\end{equation}
Here $P_b^{\mathrm{ch}}$ is obtained from
eq.~\eqref{eq:charged_response}, and
$B_{bj}$ contains the integer multiplicities
covered by experimental bin $j$.
The division by $\Delta N_{bj}$ is necessary
because the published distributions are
reported as densities rather than integrated
bin probabilities.

\begin{figure}[htbp]
\centering
\includegraphics[width=0.95\textwidth]{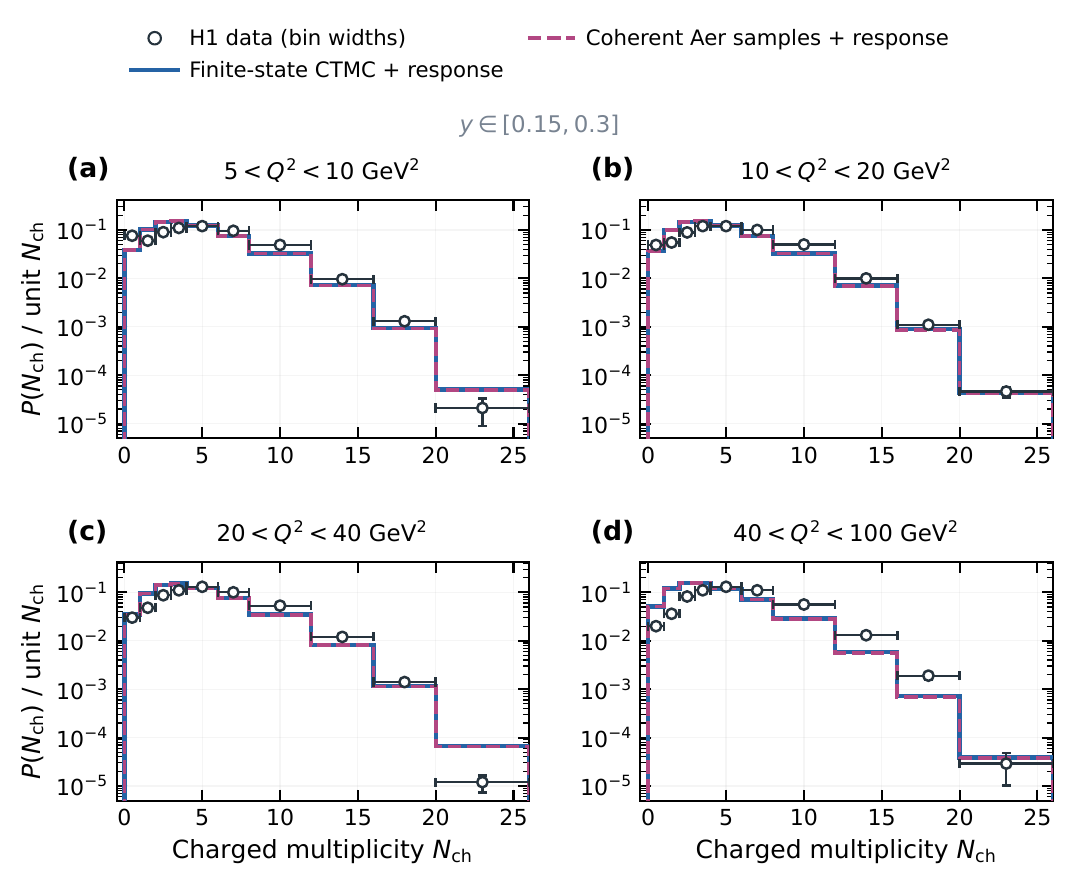}
\caption{
Charged-particle multiplicity distributions
in the hadronic centre-of-mass (HCM)
current hemisphere ($0<\eta^*<4$),
compared with H1 measurements
\cite{H1HEPData2021}.
The four panels correspond to
(a) $5<Q^2<10~\mathrm{GeV}^2$,
(b) $10<Q^2<20~\mathrm{GeV}^2$,
(c) $20<Q^2<40~\mathrm{GeV}^2$, and
(d) $40<Q^2<100~\mathrm{GeV}^2$,
for $0.15\leq y\leq0.30$.
Open circles denote the H1 data, solid
blue lines the finite-state CTMC predictions,
and dashed magenta lines the corresponding
coherent-circuit results obtained from
Qiskit Aer history measurements.
The same fixed negative-binomial response
is applied to both calculations.
The distributions are expressed as
probability densities per unit charged-particle
multiplicity, accounting for the experimental
bin widths. The logarithmic vertical scale
makes deviations in less populated
multiplicity regions visible.
}
\label{fig:h1-multiplicity}
\end{figure}

Figure~\ref{fig:h1-multiplicity} shows the
results for $0.15\leq y\leq0.30$ in the
four $Q^2$ intervals considered above.
For this binned comparison, we use the
rapidity assignment described in
appendix~\ref{app:h1-binning}.
The selected inelasticity interval corresponds
to $\Delta Y=\ln4$ relative to the reference
interval used for response calibration.
This binwise assignment differs from the
representative-$x$ mapping used for the
entropy comparison in section~\ref{subsec:dis-entropy}.

The solid curves are obtained from the
finite-state CTMC, while the dashed curves
use multiplicity probabilities sampled
from the corresponding coherent history
circuit with Qiskit Aer.
Both calculations employ the same finite
Markov generator, absorbing multiplicity
cutoff, rapidity assignment, and frozen
charged-particle response.
The Aer results are taken from the
independent history-measurement calculation,
not from the continuous-coordinate Mueller
Monte Carlo or the statevector points
shown in figure~\ref{fig:h1-entropy}.

The CTMC and Aer predictions nearly
coincide over the displayed multiplicity
range. This agreement provides a check
that sampling the coherent circuit recovers
the charged-particle distribution associated
with the prescribed finite-state Markov
process after applying the same response.
It does not constitute an independent
validation of the response model against
the H1 data.

Although the finite-state predictions
reproduce the broad structure of the
measured distributions, noticeable
differences remain. Several low-multiplicity
bins are overestimated, while the
intermediate-multiplicity region is
underestimated in parts of the measured
range. The deviations are not uniform
across the four $Q^2$ intervals, and
the less populated tail is particularly
sensitive to the distribution shape.

These differences are further illustrated
by the theory-to-data ratios in
figure~\ref{fig:app-h1-ratios}.
The ratios identify the multiplicity
regions in which the finite-state
prediction differs from the data,
but they are not a correlated
goodness-of-fit measure. Experimental
bin-to-bin correlations and uncertainties
in the phenomenological response are not
included in this diagnostic.

It is also important to distinguish
this comparison from the continuous-cascade
entropy results of the preceding subsection.
The finite circuit has a restricted
transverse configuration space and an
absorbing multiplicity boundary.
Its charged-particle distribution is
therefore a benchmark of the
finite-state implementation, rather than
a direct prediction from the larger
continuous-coordinate cascade.

Taken together, the entropy and
distribution comparisons provide
different tests of the model.
The continuous Mueller cascade,
combined with a fixed charged-particle
response, reproduces the overall increase
of the measured Shannon entropy toward
smaller $x$. The finite-state
distribution comparison demonstrates
agreement between the classical
Markov and coherent-circuit implementations,
while also showing substantial deviations
from measured distribution shapes.
The remaining discrepancies cannot be
assigned uniquely to the branching
dynamics or to the phenomenological
particle-production response.
A more complete description of the
DIS final state would require further
modeling beyond the present calculation.

\section{Conclusions and outlook}
\label{sec:conclusions}

We have constructed a coherent quantum-circuit
representation of regulated Mueller dipole branching
at small $x$. The circuit combines a coherent
Poisson clock with orthogonal event-history registers,
using local splitting rates to prepare the branching
amplitudes without loading a precomputed evolved
probability distribution. Different histories remain
distinguishable even when they reach the same final
dipole configuration, allowing the classical Markov
probabilities to be recovered from quantum measurements.
For the finite-state model, the statevector results
reproduce the truncated-clock evolution to numerical
precision and agree with an independent continuous-time
Markov-chain calculation within approximately
$7\times10^{-4}$. This establishes an explicit and
numerically verified coherent encoding of the
specified stochastic branching process.

We have also investigated the regulated Mueller
cascade through independent Monte Carlo simulations
with continuous transverse coordinates. Its
multiplicity fluctuations differ appreciably from
those of a Yule process with the same mean dipole
number. At the largest rapidity considered, the
normalized fourth factorial moment is $F_4=1.81$,
compared with $2.84$ for the Yule reference.
The Shannon and R\'enyi entropies describe the
broadening multiplicity distribution, while the
joint multiplicities of small and large dipoles
provide additional information about its transverse
structure. In particular, the size-sector covariance
continues to differ from a conditional-binomial
reference even after accounting for changes in
the mean size composition with total multiplicity.
These results show that the regulated branching
dynamics generates statistical structure that
cannot be inferred from the mean multiplicity alone.

To connect the partonic cascade with DIS measurements,
we introduced a negative-binomial response between
dipole number and charged-particle multiplicity.
The response parameters were fixed at one reference
point in each $Q^2$ interval and then held unchanged
at smaller $x$. The continuous Mueller calculation
reproduces the overall increase of the H1
charged-particle multiplicity Shannon entropy
toward smaller $x$, although quantitative
differences remain. The finite-state CTMC and
coherent-circuit sampling give consistent
charged-particle distributions under the same
response, but their distribution shapes do not
fully agree with the measurements. Thus, the
observed entropy trend provides a useful
phenomenological connection, while the detailed
multiplicity distributions expose limitations
of the present branching and particle-production
description.

The next step for the quantum algorithm is to
reduce the resources required to retain branching
histories. More economical history encodings
and gate-level decompositions would allow larger
transverse grids and longer Poisson-clock
evolutions to be investigated. The coherent
representation also provides a starting point
for measuring multiplicity characteristic functions
and, more generally, joint counting statistics
of different dipole-size sectors. Establishing
whether such measurements can offer a practical
computational advantage will require explicit
resource estimates and comparisons with classical
Markov-chain and Monte Carlo methods. This
stochastic-dynamics approach is complementary
to quantum simulations based on light-front
QCD Hamiltonians \cite{Qian:2024gph}.

On the physics side, an extension beyond the
regulated branching process would require
additional scattering dynamics, including
nonlinear small-$x$ evolution and gluon-number
fluctuation effects
\cite{Iancu:2004iy,Soyez:2005ha,
BalitskyChirilli2008}. For comparisons with
measured final states, a more detailed treatment
of particle production and experimental
acceptance would be particularly valuable.
Established event generators
\cite{Bierlich:2022pfr} and recent quantum
simulations of jet dynamics and string breaking
\cite{FlorioEtAl2024,Gonzalez-Cuadra:2024xul}
provide complementary perspectives on these
questions, although they describe different
dynamics from the present Mueller cascade.
Future measurements at the Electron-Ion
Collider may offer further opportunities to
investigate small-$x$ multiplicity fluctuations
and correlations \cite{AbdulKhalek:2021gbh}.

\acknowledgments

This work has been supported by the National Natural Science Foundation of China
(Grant No. 12547118 and 12605167) and the Research Program of State Key Laboratory of Heavy Ion
Science and Technology, Institute of Modern Physics, Chinese Academy of Sciences
(Grant No. HIST2025CS08).
\appendix

\section{Regulated dipole evolution and finite-state approximation}
\label{app:regularization}

\subsection{Transverse cutoffs and cascade evolution}
\label{app:regulated-phase-space}

We use the fixed-coupling, large-$N_c$ Mueller splitting kernel
\cite{Mueller1994Soft,MuellerPatel1994}. For a parent dipole with endpoints
$(\boldsymbol x,\boldsymbol y)$, we write its probability density per unit
rapidity and transverse area as
$\mathcal K_{\boldsymbol x\boldsymbol y}(\boldsymbol z)
=\bar\alpha_s|\boldsymbol x-\boldsymbol y|^2/
[2\pi|\boldsymbol x-\boldsymbol z|^2|\boldsymbol z-\boldsymbol y|^2]$,
consistent with eq.~\eqref{eq:mueller_kernel}. The kernel is singular as the
emission point approaches either parent endpoint. We regulate this process
by restricting the emission coordinate to the square $[-L,L]^2$ and
requiring both daughter-dipole sizes to lie between the ultraviolet and
infrared bounds:
\begin{equation}
\label{eq:app-reg-domain}
\Omega_{\mathrm{reg}}(\boldsymbol x,\boldsymbol y)
=\left\{\boldsymbol z\in[-L,L]^2:\;
\rho\leq|\boldsymbol x-\boldsymbol z|\leq R_{\mathrm{IR}},\quad
\rho\leq|\boldsymbol z-\boldsymbol y|\leq R_{\mathrm{IR}}\right\}.
\end{equation}
The transverse box limits the emission coordinate, whereas the daughter-size
conditions regulate emissions at both short and long transverse distances.
These restrictions define a finite-rate branching process, not a nonlinear
saturation or recombination mechanism \cite{Salam1995Multiplicity,Salam1997OEDIPUS}.

\begin{table}[htbp]
\centering
\caption{Physical and numerical settings of the regulated Mueller cascade.
The finite-state quantum calculation and the continuous-coordinate
Monte Carlo simulation use different multiplicity ranges.}
\label{tab:app-physical-parameters}
\begin{tabular}{lll}
\hline
Quantity & Symbol & Value \\
\hline
\multicolumn{3}{l}{\emph{Transverse dynamics}}\\
Mueller splitting coupling & $\bar\alpha_s$ & $0.20$ \\
Transverse-box half-width & $L$ & $2.0$ \\
Minimum daughter-dipole size & $\rho$ & $0.35$ \\
Maximum daughter-dipole size & $R_{\mathrm{IR}}$ & $3.5$ \\
Initial dipole-geometry parameter & $a$ & $0.50$ \\
\multicolumn{3}{l}{\emph{Numerical settings}}\\
Finite-state transverse grid & $g$ & $2$ \\
Finite-state multiplicity cap & $n_{\max}^{\mathrm{fin}}$ & $3$ \\
Continuous-cascade multiplicity cap & $n_{\max}^{\mathrm{cont}}$ & $32$ \\
Sobol integration exponent & $m_{\mathrm{Sobol}}$ & $10$ \\
Poisson-tail tolerance & $\varepsilon_{\mathrm{target}}$ & $10^{-3}$ \\
\hline
\end{tabular}
\end{table}

The initial configuration contains one dipole with endpoints
$\boldsymbol x_0=(-a,0)$ and $\boldsymbol y_0=(a,0)$, so its initial size
is $r_0=2a=1.0$ in the common model transverse-coordinate units.
If dipole $i$ emits at $\boldsymbol z$, the parent is replaced by
$(\boldsymbol x_i,\boldsymbol z)$ and $(\boldsymbol z,\boldsymbol y_i)$;
we denote the updated configuration by $\mathcal C^{(i,\boldsymbol z)}$.
The corresponding total physical splitting rate is
$\Gamma_{\mathrm{cont}}(\mathcal C)$, as defined in
\eqref{eq:total_branching_rate}.

The continuous-coordinate Monte Carlo uses an event-driven Poisson-thinning
algorithm. The transverse square is partitioned into auxiliary cells
$\Omega_j$, solely to construct upper bounds on the regulated kernel.
For a point $\boldsymbol x$, let
$d_{\min}(\boldsymbol x,\Omega_j)$ denote its minimum distance from the
cell. An admissible upper bound on the emission density in that cell is
\begin{equation}
\label{eq:app-thinning-envelope}
U_{i,j}=
\frac{\bar\alpha_s}{2\pi}
\frac{|\boldsymbol x_i-\boldsymbol y_i|^2}
{\max\{\rho^2,d_{\min}^2(\boldsymbol x_i,\Omega_j)\}
 \max\{\rho^2,d_{\min}^2(\boldsymbol y_i,\Omega_j)\}}
\end{equation}
for cells that can intersect the allowed daughter-size region; a cell
provably outside that region is assigned zero upper bound.
The proposal rate for a configuration is
$\widehat\Gamma(\mathcal C)=\sum_{i,j}|\Omega_j|U_{i,j}$.
A candidate rapidity increment is sampled from an exponential distribution
with rate $\widehat\Gamma(\mathcal C)$. The parent and proposal cell are
selected in proportion to their upper-bound rates, and a candidate
$\boldsymbol z$ is drawn uniformly within that cell. It is accepted with
probability
$\mathcal K_{\boldsymbol x_i\boldsymbol y_i}(\boldsymbol z)
\boldsymbol1_{\Omega_{\mathrm{reg}}}(\boldsymbol z)/U_{i,j}$.
A rejected candidate advances the proposal clock but leaves the physical
configuration unchanged. Consequently, accepted emissions follow the
regulated continuous-coordinate kernel: the proposal grid affects efficiency
rather than projecting the accepted $\boldsymbol z$ onto grid points.
The numerical cap and treatment of trajectories that exceed it are described
in appendix~\ref{app:mc-estimators}.

\subsection{Finite-grid evolution and multiplicity cutoff}
\label{app:finite-generator}

For the finite-state CTMC and coherent circuit, we instead partition the
transverse square into a physical set of $g\times g$ emission cells.
A channel $a=(i,j)$ specifies a parent dipole $i$ and an emission cell
$\Omega_j$. Its rate is the cell integral of the \emph{regulated} Mueller
kernel, approximated in the implementation by scrambled Sobol quadrature:
\begin{equation}
\label{eq:app-channel-rate}
\begin{aligned}
\gamma_{i,j}(\mathcal C)
&=\int_{\Omega_j\cap\Omega_{\mathrm{reg}}(\boldsymbol x_i,\boldsymbol y_i)}
 d^2\boldsymbol z\,\mathcal K_{\boldsymbol x_i\boldsymbol y_i}(\boldsymbol z)\\
&\simeq\frac{|\Omega_j|}{2^{m_{\mathrm{Sobol}}}}
\sum_{s=1}^{2^{m_{\mathrm{Sobol}}}}
\mathcal K_{\boldsymbol x_i\boldsymbol y_i}(\boldsymbol z_{js})
\boldsymbol1_{\Omega_{\mathrm{reg}}(\boldsymbol x_i,\boldsymbol y_i)}
(\boldsymbol z_{js}).
\end{aligned}
\end{equation}
Here $\boldsymbol z_{js}$ are the Sobol sample positions in cell $j$;
$m_{\mathrm{Sobol}}=10$ in the baseline finite-state calculation.
Unlike a pointwise midpoint rule, the rate is obtained from samples across
the cell. The quadrature nevertheless has finite integration resolution.

To define a finite daughter configuration, an emission cell must also
be assigned a representative transverse position. For each parent--cell
pair with positive sampled rate, let $\mathcal V_{i,j}$ be the set of
Sobol points satisfying the cutoff conditions and write
$\omega_s=\mathcal K_{\boldsymbol x_i\boldsymbol y_i}(\boldsymbol z_{js})$
for their positive weights. The kernel-weighted mean position and the
allowed representative (valid-sample medoid) are
\begin{equation}
\label{eq:app-valid-medoid}
\overline{\boldsymbol z}_{i,j}
=\frac{\sum_{s\in\mathcal V_{i,j}}\omega_s\boldsymbol z_{js}}
{\sum_{s\in\mathcal V_{i,j}}\omega_s},
\qquad
\boldsymbol z_{i,j}^{*}
=\underset{\boldsymbol z_{js}:s\in\mathcal V_{i,j}}{\operatorname{arg\,min}}
|\boldsymbol z_{js}-\overline{\boldsymbol z}_{i,j}|^2.
\end{equation}
The daughter configuration for channel $a=(i,j)$ is then
$\mathcal C_a=\mathcal C^{(i,\boldsymbol z_{i,j}^{*})}$.
This representative is selected separately for each parent geometry and
cell; it is not an unconstrained cell center. It therefore satisfies the
UV/IR daughter-size conditions for the sampled channel. This construction
approximates continuous transverse geometry and does not make the
finite-state model an exact continuous-coordinate cascade.

Let $\mathcal A(\mathcal C)$ contain all allowed channels from
$\mathcal C$ whose daughters lie in the enumerated state space
$\mathcal S$. We impose an absorbing multiplicity cap:
\begin{equation}
\label{eq:app-allowed-rate}
\Gamma(\mathcal C)=\sum_{a\in\mathcal A(\mathcal C)}\gamma_a(\mathcal C),
\qquad
\Gamma(\mathcal C)=0\quad\text{if }N(\mathcal C)=n_{\max}.
\end{equation}
With the column-probability convention used in the main text, the finite
Markov generator is
\begin{equation}
\label{eq:app-generator}
Q_{\mathcal C',\mathcal C}=
\begin{cases}
\displaystyle\sum_{a\in\mathcal A(\mathcal C):\,\mathcal C_a=\mathcal C'}
\gamma_a(\mathcal C),&\mathcal C'\ne\mathcal C,\\[5pt]
-\Gamma(\mathcal C),&\mathcal C'=\mathcal C.
\end{cases}
\end{equation}
Each column sums to zero. Equivalently, a row-stochastic generator can
be stored in the numerical implementation and transposed when it acts
on a column probability vector; both conventions represent the same CTMC.
At the absorbing boundary, further splitting is stopped, and probability
that would otherwise reach higher multiplicity accumulates in the terminal
bin. The cap is a numerical modification of the branching process, not
BK/JIMWLK saturation.

\begin{figure}[htbp]
\centering
\includegraphics[width=0.85\textwidth]{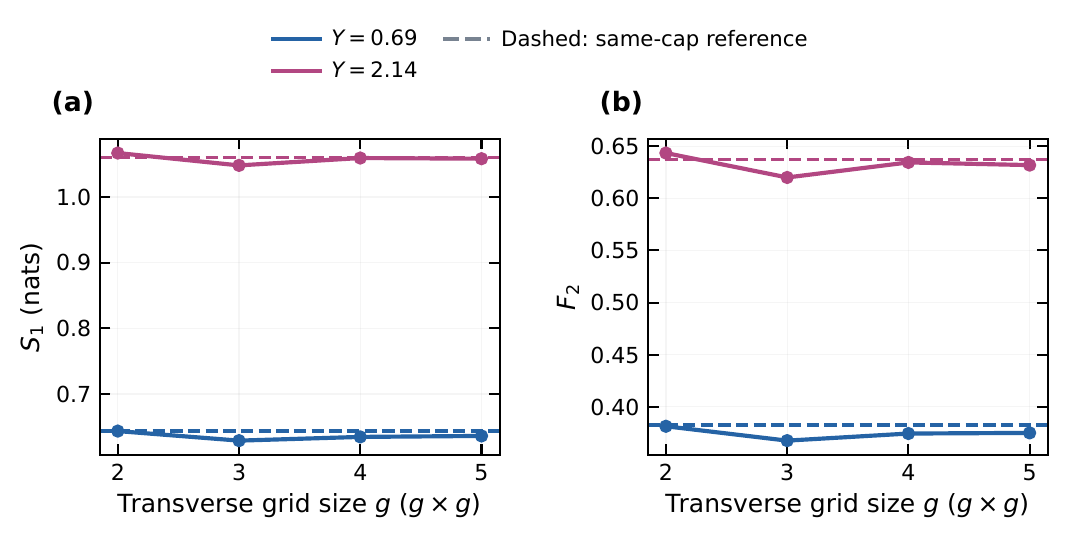}
\caption{Sensitivity of the finite-state multiplicity entropy $S_1$ and
normalized second factorial moment $F_2$ to transverse-grid resolution
$g=2,3,4,5$ at fixed multiplicity cap $n_{\max}=3$.
The two curve sets correspond to the indicated rapidities; dashed
horizontal lines denote values from a separate multiplicity-cap scan
at matching rapidity and cap. These comparisons diagnose numerical
sensitivity and do not establish convergence to the continuous-coordinate
cascade.}
\label{fig:app-grid-convergence}
\end{figure}

Figure~\ref{fig:app-grid-convergence} compares $S_1$ and $F_2$ for
$g=2,3,4,5$ at the same absorbing cap $n_{\max}=3$.
This is a finite-state transverse-grid sensitivity test at fixed cap.
It does not by itself establish a simultaneous grid-and-cap continuum limit,
nor should it be identified with changing the auxiliary envelope grid in
the continuous-coordinate Poisson-thinning algorithm.

\section{Physical interpretation of the coherent branching circuit}
\label{app:circuit-details}

\subsection{Poisson updates and the finite-clock error}
\label{app:clock}

Uniformization represents the finite-state CTMC as a Poisson-weighted
sequence of updates with a common rate
$\Lambda\geq\max_{\mathcal C}\Gamma(\mathcal C)$
\cite{Jensen1953,GrossMiller1984}. At one clock update,
configuration $\mathcal C$ either remains unchanged with probability
$1-\Gamma(\mathcal C)/\Lambda$, or splits through channel $a$ with
probability $\gamma_a(\mathcal C)/\Lambda$. These null updates are essential
because the physical splitting rates depend on the evolving configuration.
The transition matrix $T$ and Poisson probabilities $w_k(Y)$ are defined in
eq.~\eqref{eq:uniformization}.

The circuit coherently prepares the clock sectors $k=0,\ldots,K$ and an
orthogonal unresolved state $|K+1\rangle$, as in
\eqref{eq:coherent_clock_state}. The probability of this tail state is
$\varepsilon_K(Y)$, given by \eqref{eq:clock_truncation}.
The resolved multiplicity distribution, before conditioning, is
$P_{n,\mathrm{res}}^{(K)}=\sum_{k=0}^K w_k(Y)[T^k\boldsymbol p(0)]_n$.
Its sum over $n$ equals $1-\varepsilon_K(Y)$. Thus the normalized
finite-clock distribution is
\begin{equation}
\label{eq:app-clock-conditioned}
P_n^{(K)}(Y)=\frac{1}{1-\varepsilon_K(Y)}
\sum_{k=0}^{K}e^{-\Lambda Y}\frac{(\Lambda Y)^k}{k!}
[T^k\boldsymbol p(0)]_n,
\qquad
[\boldsymbol v]_n=\sum_{\mathcal C:N(\mathcal C)=n}v_{\mathcal C}.
\end{equation}
This is identical to the conditional convention of
\eqref{eq:finite_clock_conditional} after marginalization over configurations.
The physical splitting rates are not modified by conditioning; only
clock histories exceeding the retained depth are excluded.

The error bound follows directly from the Poisson mixture. Let
$P_n^{>K}(Y)$ be the properly normalized multiplicity distribution
conditional on more than $K$ ticks. Then the exact distribution is
$P_n=(1-\varepsilon_K)P_n^{(K)}+\varepsilon_K P_n^{>K}$.
Therefore
\begin{equation}
\label{eq:app-clock-bound}
\max_n|P_n^{(K)}-P_n|\leq\varepsilon_K,
\qquad
\frac12\sum_n|P_n^{(K)}-P_n|\leq\varepsilon_K.
\end{equation}
The second expression is the total-variation distance. An unconditioned
measurement of the full circuit instead yields the resolved multiplicity
probabilities plus a distinct tail-flag outcome. These two conventions
must not be mixed in comparisons with CTMC probabilities.
The bound is specific to a fixed generator and multiplicity cutoff; it
does not control transverse projection, the absorbing cap, Monte Carlo
sampling, or physical-hardware noise. If the generator has no outgoing
rates, no nontrivial clock is needed.

\subsection{Why branching histories must be retained}
\label{app:history-isometry}

Different splitting sequences can arrive at the same final configuration.
A classical Markov process adds their probabilities, whereas adding
quantum amplitudes without retaining path information could generate
unwanted interference. At the abstract level, one clock update can be
embedded in the following isometry acting on a configuration register
and a fresh history register:
\begin{align}
\label{eq:app-isometry}
V:\ |\mathcal C\rangle|0\rangle_h\longmapsto {}&
\sqrt{1-\frac{\Gamma(\mathcal C)}{\Lambda}}
|\mathcal C\rangle|\mathcal C,0\rangle_h\\
&+\sum_{a\in\mathcal A(\mathcal C)}
\sqrt{\frac{\gamma_a(\mathcal C)}{\Lambda}}
|\mathcal C_a\rangle|\mathcal C,a\rangle_h.\nonumber
\end{align}
The first term represents a null update and the others the physical
splitting channels. The output history labels include the source
configuration and channel; they are orthogonal for distinct inputs and
outcomes. Hence $V^\dagger V=I$ on the input subspace, and the isometry
can be extended to a unitary operation on a sufficiently large space.

Equation~\eqref{eq:app-isometry} is a convenient \emph{conceptual}
construction, not the literal register allocation used in our Qiskit
state preparation. For the fixed initial dipole $\mathcal C_0$, the actual
circuit stores the clock value and an ordered sequence of event codes
$h=(e_1,\ldots,e_k)$, with $e_t=0$ for a null update. The current
configuration $\mathcal C_t(h)$ is determined recursively from the initial
dipole and the preceding codes. Its one-step probabilities are the local
$\pi(e\mid\mathcal C)$ of eq.~\eqref{eq:local_event_probabilities}.
Controlled rotations encode these probabilities conditionally on the
clock and history prefix. The resulting amplitude for a length-$k$
history is $\sqrt{w_k(Y)\mathbb P(h\mid k)}$, giving exactly the state
in eq.~\eqref{eq:coherent_history_state}. The construction uses local
transition information rather than a precomputed evolved distribution.

Histories, including null-update labels, are retained in orthogonal
basis states. The final configuration or multiplicity is reconstructed
from those labels, and outcome probabilities are \emph{grouped} by the
reconstructed observable. This gives
$[T^k\boldsymbol p(0)]_{\mathcal C}$, as in
\eqref{eq:coherent_probability}, without summing indistinguishable
probability amplitudes. The representation is coherent in the quantum
register space but encodes a classical stochastic process; it does not
model interference between QCD emission amplitudes or real-time
evolution under the QCD Hamiltonian.

\subsection{What the finite circuit measures}
\label{app:resources-fcs}

The implemented encoding uses one register for the Poisson clock and
$K$ event-history registers. For a $g\times g$ emission grid and an
absorbing cap $n_{\max}$, each active update has one null-event code
and at most $(n_{\max}-1)g^2$ splitting-channel codes. The logical
qubit count in the implemented state-preparation circuit is therefore
\begin{equation}
\label{eq:app-resource-accounting}
\begin{aligned}
n_{\mathrm{clock}}&=\left\lceil\log_2(K+2)\right\rceil,\\
n_{\mathrm{event}}&=\left\lceil\log_2[1+(n_{\max}-1)g^2]\right\rceil,\\
n_{\mathrm{qubits}}&=n_{\mathrm{clock}}+K\,n_{\mathrm{event}}.
\end{aligned}
\end{equation}
No additional quantum register for the full configuration or
multiplicity is included in this count: the state is recovered from
the event history. An ancilla-controlled characteristic-function
measurement adds one extra logical qubit. The above count does not
include hardware error correction, physical-qubit overhead, or a
standard-gate decomposition of multiply controlled rotations.

\begin{figure}[htbp]
\centering
\includegraphics[width=0.86\textwidth]{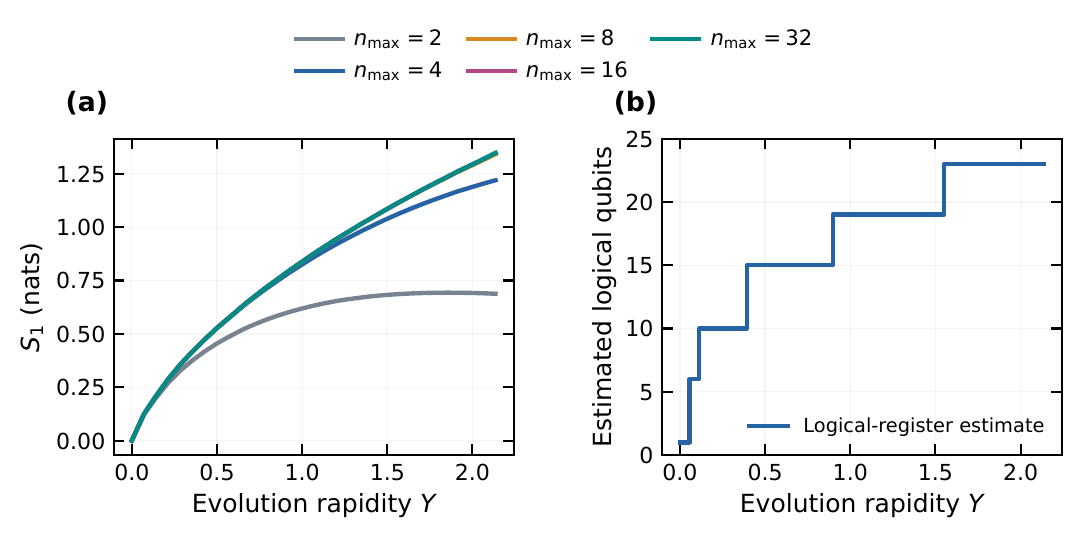}
\caption{Distinct limitations from the multiplicity cutoff and
coherent event-history storage.
(a) Shannon entropy $S_1(Y)$ for different absorbing multiplicity
caps $n_{\max}$.
(b) Logical-qubit requirements estimated for the clock-and-history
encoding at successive retained clock depths. Configurations and
multiplicities are reconstructed from events rather than stored in
separate registers. The values are encoding estimates, not counts
of physical hardware qubits or compiled native gates.}
\label{fig:app-cap-resources}
\end{figure}

For the baseline $g=2$ and $n_{\max}=3$ model, the event code uses
four qubits. The implemented $K=3$ and $K=4$ statevector circuits
therefore require 15 and 19 logical qubits, respectively. Their
construction contains 148 and 333 controlled $R_y$ rotations before
hardware-native decomposition. A $K=5$ extension would require 23
logical qubits in the same encoding; this is a resource estimate,
not a completed 23-qubit statevector run. Figure~\ref{fig:app-cap-resources}
separates the sensitivity to the absorbing cap from the stepwise
history-storage cost.

The final dipole number associated with a measured history is
$N(h)=1+\sum_t\boldsymbol1_{e_t\ne0}$, as in
\eqref{eq:history_multiplicity}; grouping clock-and-history outcomes
recovers $P_{n,\mathrm{res}}^{(K)}$. The multiplicity characteristic
function is defined by eq.~\eqref{eq:multiplicity_characteristic}.
For the finite-depth circuit, we distinguish the resolved Fourier sum
$\chi_{\mathrm{res}}^{(K)}$ from the normalized conditional result
$\chi^{(K)}=\chi_{\mathrm{res}}^{(K)}/(1-\varepsilon_K)$.
When the unresolved clock flag receives zero counting phase, the raw
ancilla-controlled signal is
\begin{equation}
\label{eq:app-ancilla-tail}
\langle X\rangle+i\langle Y\rangle
=\varepsilon_K(Y)+\chi_{\mathrm{res}}^{(K)}(\phi,Y),
\qquad
\chi^{(K)}(\phi,Y)
=\frac{\langle X\rangle+i\langle Y\rangle-\varepsilon_K(Y)}
{1-\varepsilon_K(Y)}.
\end{equation}
The phase operator assigns $e^{i\phi}$ to the initial resolved dipole
and one additional factor for every nonzero event code; the tail-flag
sector carries phase one. This implements the dipole-counting phase
without a separate multiplicity register. The circuit uses controlled
rotations for recursive probability-state preparation, a general
technique discussed in refs.~\cite{GroverRudolph2002,Mottonen2005StatePrep}.

\begin{figure}[htbp]
\centering
\includegraphics[width=0.84\textwidth]{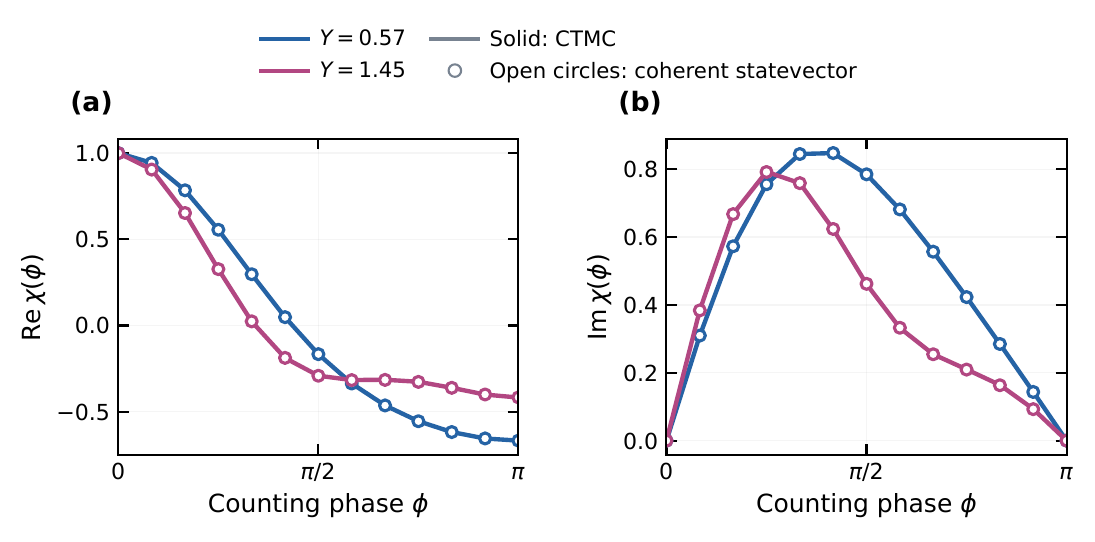}
\caption{Real and imaginary parts of the multiplicity characteristic
function $\chi(\phi,Y)$ obtained from the finite coherent statevector
and the matched CTMC distribution at two rapidities.
The statevector values are derived from the coherent numerical state;
this figure does not represent independent ancilla sampling or a
quantum-hardware experiment.}
\label{fig:app-characteristic}
\end{figure}

Figure~\ref{fig:app-characteristic} compares the real and imaginary
parts obtained \emph{directly from statevectors} with the matched
finite-state CTMC; it is not a plot of independently sampled Hadamard
measurements. Separately, Qiskit Aer ancilla measurements in the $X$
and $Y$ bases have been performed at the selected phase $\phi=0.7$
with 8192 shots per basis. They agree with the corresponding
statevector expectation values within sampling precision. These are
classically simulated circuit measurements, not quantum-hardware data.

\section{Statistical tests of the transverse cascade}
\label{app:statistics}

\subsection{Event sampling and higher moments}
\label{app:mc-estimators}

The continuous-coordinate results are obtained from $M=10^5$
independent cascade trajectories with maximum dipole multiplicity
$n_{\max}=32$, evaluated at 41 rapidity values. The trajectory
sample is assembled from eight independently seeded batches. The
same trajectories are followed across rapidities, so results at
different $Y$ are statistically correlated even though the event
histories are independently sampled from one trajectory to another.

When a physical splitting would take a trajectory beyond the cap,
the event is recorded as \emph{overflow}; its probability is not
added to the terminal multiplicity bin. For retained multiplicities
$n=1,\ldots,n_{\max}$, the unconditional empirical probabilities and
their same-rapidity multinomial covariance are
\begin{equation}
\label{eq:app-mc-prob}
\widehat P_n^{\mathrm{uncond}}(Y)
=\frac1M\sum_{m=1}^M\boldsymbol1_{\{n_m(Y)=n\}},
\qquad
\operatorname{Cov}(\widehat P_n,\widehat P_{n'})
=\frac{P_n\delta_{nn'}-P_nP_{n'}}{M}.
\end{equation}
The multinomial categories include an additional overflow outcome.
Writing $M_{\mathrm{ret}}=M-M_{\mathrm{over}}$, the conditional
probabilities used for the factorial moments and information entropies
are
\begin{equation}
\label{eq:app-mc-conditional}
\widehat P_n^{\mathrm{cond}}(Y)
=\frac{\#\{m:n_m(Y)=n\}}{M_{\mathrm{ret}}}
=\frac{\widehat P_n^{\mathrm{uncond}}(Y)}
{1-\widehat P_{\mathrm{over}}(Y)}.
\end{equation}
This conditioning is necessary whenever overflow is nonzero and
must not be presented as an unbiased estimate of the unrestricted
infinite-multiplicity process. For the reported $10^5$-trajectory
sample, zero overflow events are observed throughout the evaluated
rapidity range. The one-sided uncertainty therefore remains finite;
for example, the 95\% Wilson upper limit for a zero count among
$10^5$ trials is approximately $3.8\times10^{-5}$.

The probability error bars use 95\% Wilson binomial intervals for
individual multiplicity categories. For the mean, $F_2,F_3,F_4$ and
$S_1,S_2,S_3$, 95\% percentile intervals are obtained by multinomial
bootstrap resampling of all $M$ outcomes, \emph{including} the overflow
category, followed by conditional normalization for each resample.
The reported analysis uses 400 bootstrap resamples per rapidity.
In particular, high-order moments weight rare large-$n$ events heavily;
their uncertainty must be assessed together with overflow and cap
sensitivity rather than inferred from the mean multiplicity alone.
These entropies are classical information entropies of the sampled
multiplicity probabilities.

\subsection{Why compare with an equal-mean Yule process?}
\label{app:yule-cap}

In the Yule pure-birth process, each existing dipole splits at the
same rate $\lambda$. Its total rate is $n\lambda$, independent of
transverse geometry. For one initial dipole its multiplicity
probability is geometric:
\begin{equation}
\label{eq:app-yule-dist}
P_n^{\mathrm{Yule}}(Y)=p(1-p)^{n-1},
\qquad
p=e^{-\lambda Y}=1/\langle n\rangle_Y^{\mathrm{Yule}},
\qquad n\geq1.
\end{equation}
At each rapidity we choose
$p(Y)=1/\langle n\rangle_Y^{\mathrm{Mueller}}$.
This matches the mean at that point; it does not assert the existence
of a single fixed $\lambda$ reproducing the entire Mueller rapidity
evolution. With the mean so matched, the exact Yule moments and
entropies are
\begin{align}
\label{eq:app-yule-moments}
F_q^{\mathrm{Yule}}(Y)
&=q!\left(1-\frac1{\langle n\rangle_Y^{\mathrm{Mueller}}}\right)^{q-1},\\
\label{eq:app-yule-entropies}
S_1^{\mathrm{Yule}}&=-\ln p-\frac{1-p}{p}\ln(1-p),\nonumber\\
S_\alpha^{\mathrm{Yule}}&=
\frac{\alpha\ln p-\ln[1-(1-p)^\alpha]}{1-\alpha}.
\end{align}
These expressions are understood by continuity at the single-dipole
initial state $p=1$. They provide an analytic reference for the \emph{shape}
of $P_n$, without attributing observed differences to a unique
microscopic correlation mechanism.

\begin{figure}[htbp]
\centering
\includegraphics[width=0.84\textwidth]{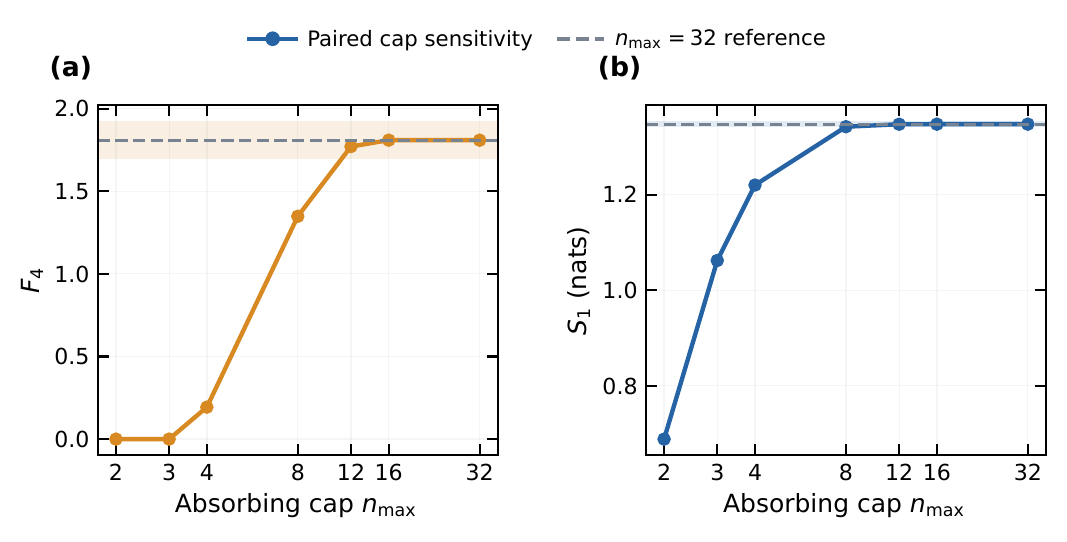}
\caption{Sensitivity of (a) the fourth normalized factorial moment
$F_4$ and (b) Shannon entropy $S_1$ to an absorbing multiplicity cap,
evaluated using paired cascade trajectories. Dashed horizontal lines
indicate the $n_{\max}=32$ reference results; the corresponding shaded
bands show their 95\% Monte Carlo confidence intervals. The cap scan
characterizes finite-multiplicity effects within the regulated cascade.}
\label{fig:app-cap-convergence}
\end{figure}

The fourth factorial moment is particularly sensitive to the
large-multiplicity tail. To examine the effect of an absorbing cap
within the same regulated continuous-coordinate dynamics, we use
paired trajectory histories and aggregate the probability reaching
$n\geq n_{\max}$ into the terminal bin for each trial cap.
This \emph{absorbing-cap diagnostic} differs from the separate
overflow category of the primary $n_{\max}=32$ data.
Figure~\ref{fig:app-cap-convergence} compares $F_4$ and $S_1$ under
these cap choices. It assesses stability within the displayed
rapidity interval, not convergence at arbitrarily large rapidity.
The underlying cap identities are derived in appendix~\ref{app:cap-aggregation}.

\subsection{Correlations between small and large dipoles}
\label{app:joint-size}

We divide dipoles at $r_{\mathrm{cut}}=0.50$ into small and large
sectors with counts $n_s$ and $n_l$, respectively, where $n_s+n_l=n$.
A positive unconditional covariance need not imply a positive
correlation at fixed $n$: fluctuations in the total number of dipoles
can increase both sector populations together. The primary
fixed-fraction reference preserves $P_n(Y)$ but assigns each dipole
to the small sector with the same probability
$p_s(Y)=\langle n_s\rangle_Y/\langle n\rangle_Y$.
Conditional on $n$, $n_s\sim\operatorname{Binomial}(n,p_s)$, yielding
\begin{equation}
\label{eq:app-binomial-cov}
C_{sl}^{\mathrm{bin}}(Y)
=p_s(Y)[1-p_s(Y)]
[\operatorname{Var}(n)_Y-\langle n\rangle_Y].
\end{equation}
This is the reference displayed in figure~\ref{fig:size-correlations}.
It does not preserve the measured variation of the mean size composition
with total multiplicity.

We therefore introduce a conditional composition reference using
$p_s(n,Y)=\langle n_s\mid n,Y\rangle/n$. Denote the measured
total-multiplicity probability by $P_n(Y)$ and let
$f_n=np_s(n,Y)$ and $g_n=n[1-p_s(n,Y)]$.
In this reference the conditional mean small- and large-dipole numbers
are $f_n$ and $g_n$, while the conditional small-dipole variance is
assumed to be $np_s(n,Y)[1-p_s(n,Y)]$. The law of total covariance
gives
\begin{equation}
\label{eq:app-condbin-cov}
\begin{aligned}
C_{sl}^{\mathrm{bin},n}(Y)
&=\operatorname{Cov}_{P_n}(f_n,g_n)
-\sum_nP_n(Y)\,n p_s(n,Y)[1-p_s(n,Y)]\\
&=\sum_nP_n(Y)f_ng_n
-\Bigl(\sum_nP_n(Y)f_n\Bigr)
 \Bigl(\sum_nP_n(Y)g_n\Bigr)
-\sum_nP_n(Y)\,n p_s(n,Y)[1-p_s(n,Y)].
\end{aligned}
\end{equation}
At $n=1$ the initial dipole has a definite size category, so its
conditional $p_s(1,Y)$ is either zero or one, as required by the
measured joint distribution.

For the true conditional distribution at fixed $n$, the identity
$n_l=n-n_s$ implies
$\operatorname{Cov}(n_s,n_l\mid n,Y)
=-\operatorname{Var}(n_s\mid n,Y)\leq0$.
Subtracting the conditional-binomial reference therefore yields
\begin{equation}
\label{eq:app-excess-cov-identity}
\Delta C_{sl}(Y)
=C_{sl}(Y)-C_{sl}^{\mathrm{bin},n}(Y)
=\sum_nP_n(Y)
\left\{n p_s(n,Y)[1-p_s(n,Y)]
-\operatorname{Var}(n_s\mid n,Y)\right\}.
\end{equation}
A positive $\Delta C_{sl}$ consequently means that the measured
size-composition fluctuations, averaged over $n$, are \emph{smaller}
than for independent binomial assignments at the same conditional
means. It does not imply positive covariance at fixed multiplicity.

This diagnostic uses $6\times10^4$ independent continuous-coordinate
trajectories, rather than the $10^5$ trajectories used for the total
multiplicity study. At $Y=2.1355$, we find
$C_{sl}=0.07096$, $C_{sl}^{\mathrm{bin}}=-0.07677$, and
$C_{sl}^{\mathrm{bin},n}=0.02975$. The excess above the refined
reference is $\Delta C_{sl}=0.04120$, with a 95\% bootstrap confidence
interval $[0.03748,0.04505]$. The interval is obtained by resampling
the joint multiplicity counts and recomputing the conditional
reference for every resample. This is a statistical diagnostic of
within-$n$ size composition, not evidence for quantum entanglement
or a unique microscopic origin of the residual deviation.

\section{DIS response and comparison with H1}
\label{app:dis-response}

\subsection{From dipole counts to charged particles}
\label{app:nbd-response}

The dipole number $n$ is not directly the observed charged-particle
multiplicity $N_{\mathrm{ch}}$. We introduce the conditional response
$R_b(N\mid n;\boldsymbol\vartheta_b)$ of
\eqref{eq:charged_response}, where $b$ denotes one of the four
$Q^2$ intervals. In each interval the response is a negative-binomial
distribution with the mean $\mu_b(n)=n\mu_b^{(0)}$ and shape
$\kappa_b(n)=nr_b^{(0)}$:
\begin{equation}
\label{eq:app-nbd}
R_b(N\mid n;\boldsymbol\vartheta_b)
=\frac{\Gamma_{\!E}(N+\kappa_b(n))}
{\Gamma_{\!E}(\kappa_b(n))\Gamma_{\!E}(N+1)}
\left(\frac{\kappa_b(n)}{\kappa_b(n)+\mu_b(n)}\right)^{\kappa_b(n)}
\left(\frac{\mu_b(n)}{\kappa_b(n)+\mu_b(n)}\right)^N,
\qquad N=0,1,\ldots\;.
\end{equation}
Here $\Gamma_{\!E}$ denotes Euler's gamma function, to distinguish it
from the dipole splitting rate $\Gamma(\mathcal C)$.
The assumed independent, identically parameterized particle-production
sources add their negative-binomial means and shape parameters.
This is a phenomenological response, not a microscopic hadronization
calculation.

For the stated mean--shape parameterization, the single-source variance
is $\mu_b^{(0)}+[\mu_b^{(0)}]^2/r_b^{(0)}$.
Since the initial condition at $Y=0$ consists of one dipole, the
response parameters can be fixed from the published reference-point
mean $\overline N_{b,0}$ and variance $V_{b,0}$:
\begin{equation}
\label{eq:app-nbd-calibration}
\mu_b^{(0)}=\overline N_{b,0},
\qquad
r_b^{(0)}=
\frac{\overline N_{b,0}^{\,2}}{V_{b,0}-\overline N_{b,0}},
\qquad V_{b,0}>\overline N_{b,0}.
\end{equation}
No entropy observable is directly fitted. The fitted parameters are
subsequently held fixed at every other $x$ point in the same $Q^2$
interval. The response-convoluted multiplicity has mean and variance
\begin{equation}
\label{eq:app-response-moments}
\begin{aligned}
\langle N_{\mathrm{ch}}\rangle_Y
&=\mu_b^{(0)}\langle n\rangle_Y,\\
\operatorname{Var}(N_{\mathrm{ch}})_Y
&=\left[\mu_b^{(0)}+
\frac{[\mu_b^{(0)}]^2}{r_b^{(0)}}\right]\langle n\rangle_Y
+[\mu_b^{(0)}]^2\operatorname{Var}(n)_Y.
\end{aligned}
\end{equation}
The second term demonstrates explicitly that partonic dipole-number
fluctuations contribute to the observed charged-particle variance,
but only through the assumed response.

\begin{table}[htbp]
\centering
\caption{Fixed single-dipole negative-binomial response parameters and
rapidity anchors for the four photon-virtuality intervals.
The response parameters are determined from the reference-point
charged-particle mean and variance and held fixed as $x$ varies.}
\label{tab:app-frozen-response}
\begin{tabular}{cccc}
\hline
$Q^2$ interval ($\mathrm{GeV}^2$) & $\mu_b^{(0)}$ & $r_b^{(0)}$ & $x_{\mathrm{ref},b}$ \\
\hline
$5<Q^2<10$ & $3.00$ & $18.00$ & $0.0014$ \\
$10<Q^2<20$ & $3.00$ & $22.50$ & $0.0023$ \\
$20<Q^2<40$ & $3.10$ & $19.22$ & $0.0052$ \\
$40<Q^2<100$ & $2.80$ & $11.20$ & $0.0110$ \\
\hline
\end{tabular}
\end{table}

For the entropy comparison we use the rapidity convention
\begin{equation}
\label{eq:app-rapidity-map}
Y_b(x)=\ln\frac{x_{\mathrm{ref},b}}{x},
\qquad x\leq x_{\mathrm{ref},b}.
\end{equation}
There are 16 H1 entropy points across the four $Q^2$ intervals:
four reference points constrain the response and the remaining 12
provide comparisons at smaller $x$ without further fitting.
The reference points are not independent predictions of rapidity
evolution. The charged-particle response is evaluated over
$0\leq N_{\mathrm{ch}}\leq240$; the omitted tail of each used
negative-binomial response column is checked to be below $10^{-9}$.
The column tails are not silently recovered by an extra
renormalization. The Shannon entropy of the resulting distribution is
a classical count entropy, not quantum entanglement entropy.

\subsection{Experimental acceptance and distribution shapes}
\label{app:h1-binning}

We use the H1 charged-particle multiplicity measurements in the
current hemisphere $0<\eta^*<4$ of the hadronic centre-of-mass
(HCM) frame \cite{H1HEPData2021}. The selected charged tracks also
satisfy $|\eta_{\mathrm{lab}}|<1.6$. The multiplicity entropies and
distributions are taken from table~8 and tables~25--28,
respectively, of ref.~\cite{H1HEPData2021}.
Four photon-virtuality intervals are used:
$5<Q^2<10$, $10<Q^2<20$, $20<Q^2<40$, and
$40<Q^2<100~\mathrm{GeV}^2$. The representative entropy
comparison uses the 16 published $x$ values and the mapping in
\eqref{eq:app-rapidity-map} rather than integrating the theoretical
prediction over each experimental acceptance bin.

\begin{figure}[htbp]
\centering
\includegraphics[width=0.90\textwidth]{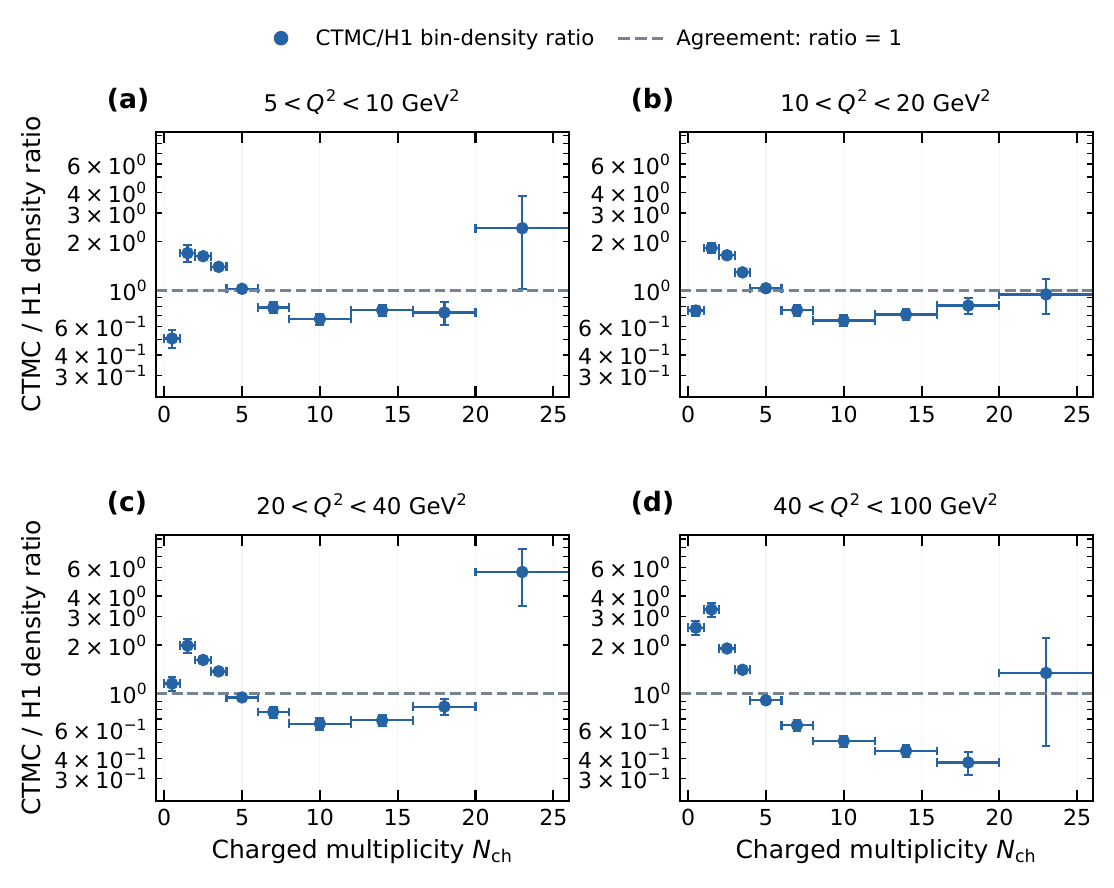}
\caption{Ratios of finite-state CTMC predictions to H1
charged-particle multiplicity densities in the HCM current hemisphere
($0<\eta^*<4$), evaluated for the four $Q^2$ intervals.
All densities account for the experimental multiplicity-bin widths.
The dashed horizontal line indicates unit theory-to-data ratio;
horizontal error bars denote bin half-widths and vertical error bars
propagate displayed H1 uncertainties in an uncorrelated approximation.
Neither correlated experimental uncertainties nor response-model
uncertainties are included in these illustrative ratios.}
\label{fig:app-h1-ratios}
\end{figure}

For the distribution comparison, the experimental bins have finite,
sometimes unequal, widths. The predicted density is obtained by
summing the integer-multiplicity probabilities over each bin and
dividing by the width, as in eq.~\eqref{eq:h1_bin_density}.
The Shannon entropy is defined from probabilities and is not
obtained by inserting bin-width-normalized densities into its
discrete entropy formula. For the binned distribution comparison,
the inelasticity intervals
$0.0375<y<0.075$, $0.075<y<0.15$, and $0.15<y<0.30$
are assigned rapidity increments $\Delta Y=0$, $\ln2$, and $\ln4$,
respectively. The first interval supplies the response calibration;
the latter intervals use the same frozen parameters.
Figures~\ref{fig:h1-multiplicity} and~\ref{fig:app-h1-ratios}
show the highest-$y$ interval. This discrete binwise prescription
and the representative-$x$ entropy mapping are different
approximations to experimental kinematics; they should not be
identified as the same phase-space integration.

Experimental bins span nonzero ranges of $Q^2$ and $y$, with
$x=Q^2/(sy)$ in DIS kinematics. Results evaluated at a representative
$x$ or assigned a single $\Delta Y$ do not constitute an
acceptance-integrated event-generator prediction.
The response likewise introduces model dependence beyond the
regulated partonic cascade. The distribution-level comparison in
figure~\ref{fig:h1-multiplicity} uses the finite-state CTMC
and corresponding Qiskit Aer history samples; the
continuous-coordinate Monte Carlo provides the primary entropy
comparison in figure~\ref{fig:h1-entropy}. Those distinct
calculations share a response prescription but not an identical
finite-state generator or multiplicity cutoff.

To display where the finite-state distribution differs from the
experimental density, we define
\begin{equation}
\label{eq:app-h1-ratio}
\mathcal R_{bj}
=\frac{D_{bj}^{\mathrm{th}}}{D_{bj}^{\mathrm{H1}}},
\qquad D_{bj}^{\mathrm{H1}}>0.
\end{equation}
This ratio is a descriptive shape diagnostic, not a correlated
goodness-of-fit statistic. It does not include the full experimental
bin-to-bin covariance or uncertainty in the phenomenological response.

\begin{figure}[htbp]
\centering
\includegraphics[width=0.90\textwidth]{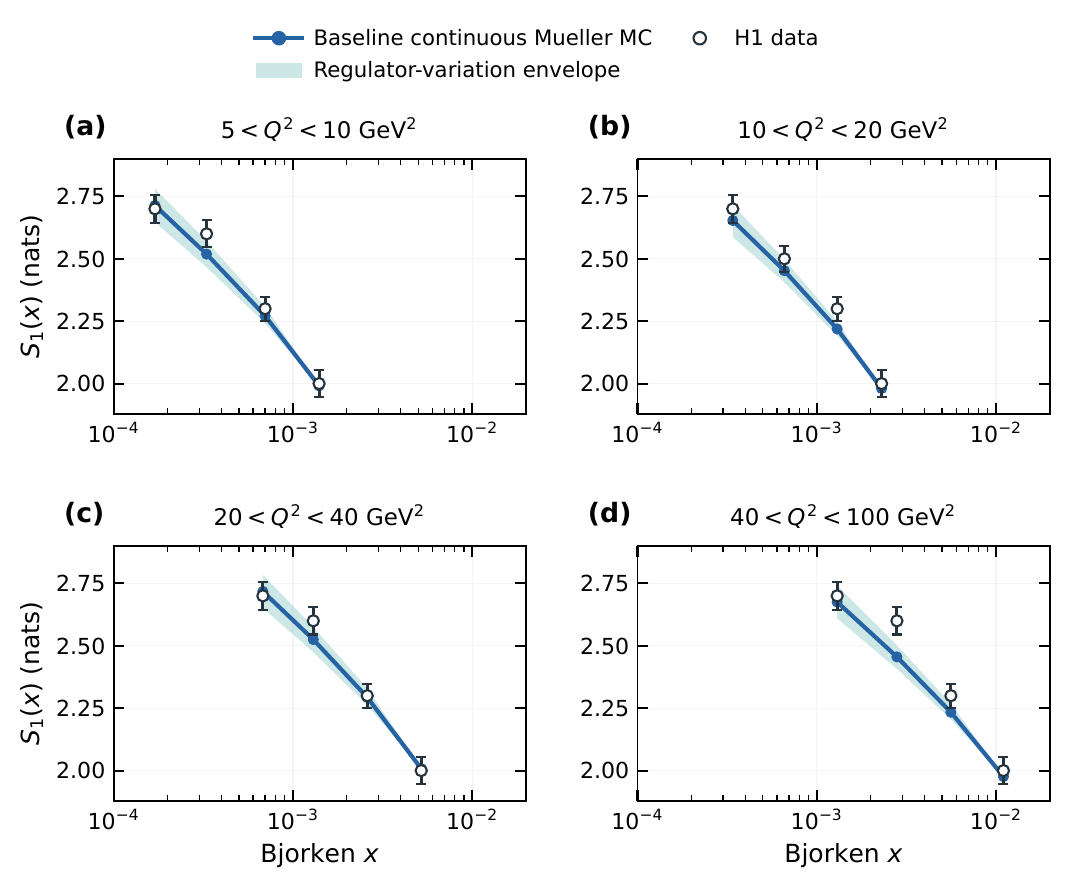}
\caption{Sensitivity of the H1 charged-multiplicity Shannon-entropy
comparison in the HCM current hemisphere ($0<\eta^*<4$) to variations
of the transverse regulators. Curves show the baseline continuous
Mueller Monte Carlo predictions after the fixed charged-particle
response; the shaded envelope spans the regulator variants
considered, not a statistical confidence interval.
Experimental points are the published H1 measurements.}
\label{fig:app-h1-regulators}
\end{figure}

We also test the sensitivity of the entropy comparison to the
physical transverse regulators, varying one at a time by $\pm10\%$
about $\rho=0.35$ and $R_{\mathrm{IR}}=3.5$.
The four variants are $\rho=0.315,0.385$ with
$R_{\mathrm{IR}}=3.5$ fixed, and $R_{\mathrm{IR}}=3.15,3.85$
with $\rho=0.35$ fixed. Other physical parameters and the
charged-particle response remain fixed. The envelope plotted in
figure~\ref{fig:app-h1-regulators} spans the baseline prediction
and these four model variants. Since changing the regulator changes
the \emph{regulated physical process}, this envelope is not a
statistical confidence interval or an estimator convergence band.
The theory-to-data ratios for the finite-state distribution are
shown separately in figure~\ref{fig:app-h1-ratios}.

\section{Analytic checks and the physical effect of the multiplicity cap}
\label{app:analytic-validation}

\subsection{Zero, one, and two successive emissions}
\label{app:small-rapidity-checks}

The first few physical splitting sectors give analytic checks of
both the finite-state Markov generator and its coherent encoding.
These sectors count \emph{accepted physical emissions}, not null
Poisson-clock updates. Let the initial dipole have total outgoing
rate $\Gamma_0$. The probability of no physical emission by rapidity
$Y$ is exactly
\begin{equation}
\label{eq:app-exact-one}
P_1(Y)=e^{-\Gamma_0Y}.
\end{equation}
For a first physical splitting through channel $a$ with rate
$\gamma_a$, let the daughter configuration have total outgoing
rate $\Gamma_a$. Exactly two dipoles are present at rapidity $Y$
when the first splitting occurs and no subsequent splitting occurs.
For $n_{\max}\geq3$, integrating over the first-splitting time yields
\begin{equation}
\label{eq:app-exact-two}
P_2(Y)=\sum_a\gamma_a
\frac{e^{-\Gamma_aY}-e^{-\Gamma_0Y}}{\Gamma_0-\Gamma_a}.
\end{equation}
The equal-rate limit $\Gamma_a=\Gamma_0$ is
$Y e^{-\Gamma_0Y}$ for the corresponding term.
For an absorbing cap $n_{\max}=2$, all two-dipole configurations
are terminal, and instead $P_2(Y)=1-e^{-\Gamma_0Y}$.

At small rapidity, zero, one and two emissions contribute at
orders $Y^0$, $Y^1$ and $Y^2$, respectively:
\begin{align}
\label{eq:app-low-Y-series}
P_1(Y)&=1-\Gamma_0Y+\tfrac12\Gamma_0^2Y^2+O(Y^3),\\
P_2(Y)&=\Gamma_0Y-\tfrac12\left(\Gamma_0^2+
\sum_a\gamma_a\Gamma_a\right)Y^2+O(Y^3),\nonumber\\
P_3(Y)&=\tfrac12\sum_a\gamma_a\Gamma_a\,Y^2+O(Y^3).\nonumber
\end{align}
These expansions assume the model permits two successive physical
splittings, as in the baseline cap $n_{\max}=3$.
The $Y^2$ coefficient of $P_3$ probes both the initial rates and
the outgoing rates of the resulting daughter configurations.
The identities are independent of the particular Poisson-clock depth
in the exact $K\to\infty$ limit; for finite $K$ they provide
corresponding checks after the controlled clock truncation.

\subsection{What an absorbing multiplicity cap changes}
\label{app:cap-aggregation}

For a pure-birth process with nondecreasing multiplicity, stopping
branching when a trajectory first reaches $n_{\max}$ preserves
the probabilities at all smaller multiplicities but aggregates
the probability of larger multiplicities in the terminal bin.
For otherwise identical dynamics, this gives
\begin{equation}
\label{eq:app-cap-tail-identity}
P_n^{\mathrm{cap}}(Y)=P_n(Y)\quad(n<n_{\max}),
\qquad
P_{n_{\max}}^{\mathrm{cap}}(Y)
=\sum_{n\geq n_{\max}}P_n(Y).
\end{equation}
This exact aggregation identity applies to a matched absorbing-cap
construction. It should not be confused with the primary continuous
Monte Carlo analysis, where trajectories exceeding $n_{\max}=32$
are counted in a separate overflow category and retained moments
are conditioned on no overflow.

For the unnormalized factorial moment, the decrease caused by
terminal-bin aggregation is nonnegative:
\begin{equation}
\label{eq:app-cap-factorial-bias}
\langle(N)_q\rangle
-\langle(N_{\mathrm{cap}})_q\rangle
=\sum_{n>n_{\max}}P_n(Y)
\bigl[(n)_q-(n_{\max})_q\bigr]\geq0.
\end{equation}
However, the normalized factorial moment
$F_q=\langle(N)_q\rangle/\langle N\rangle^q$
has no general monotonicity under the cap, because both its
numerator and mean-multiplicity denominator change.
The mapping $N\mapsto\min(N,n_{\max})$ is deterministic and
coarse-grains the original count distribution; hence its Shannon
entropy cannot exceed the entropy of that same uncapped distribution.
These facts explain the need for separate cap checks when
interpreting higher-order moments and entropies. They apply only
to matched pure-birth dynamics and do not equate the finite-grid
circuit benchmark with the independently sampled continuous-coordinate
cascade.

\bibliographystyle{JHEP}
\bibliography{ref}

\end{document}